\documentclass[]{article}
\usepackage{graphicx}
\usepackage{color}
\usepackage{rotating}

\title{Single crystal growth from light, volatile, and reactive materials using lithium and calcium flux.}
\author{A. Jesche$^1$ and P. C. Canfield$^{1,2}$} 

\begin{document}
\maketitle
$^1$\textit{The Ames Laboratory, Iowa State University, Ames, Iowa, USA}\\
$^2$\textit{Department of Physics and Astronomy, Iowa State University, Ames, Iowa, USA}

\begin{abstract}
\noindent
We present a method for the solution growth of single crystals from reactive Li and Ca melts and its application to the synthesis of several, representative compounds.
Among these, single crystalline Li$_3$N, Li$_2$(Li$_{1-x}T_x$)N with $T$ = \{Mn, Fe, Co\}, LiCaN, Li$_2$C$_2$, LiRh, and LiIr from Li-rich flux as well as Ca$_2$N, CaNi$_2$, CaNi$_3$, YbNi$_2$, Y$_2$Ni$_7$, and LaNi$_5$ from Ca-rich flux could be obtained.
Special emphasize is given on the growth of nitrides using commercially available Li$_3$N and Ca$_3$N$_2$ powders as the nitrogen source instead of N$_2$ gas. 
\end{abstract}

\tableofcontents
\section{Introduction}
Single crystals allow for direct access to the anisotropy of physical properties.
Furthermore, the characterization of single crystalline material by means of electric and thermal transport measurements often reflects intrinsic properties to a much higher extent since measurements on polycrystals can be severely affected by grain boundaries, texture, and strain.
\\
For reactive materials there is another advantage: single crystals can be significantly less air-sensitive than polycrystalline or powdered samples due to the formation of a passivating surface layer. 
This is of great practical value for a basic characterization of magnetic as well as electrical and thermal transport properties. As an example, sample mounting and the transfer of samples from a glovebox to a specific experimental setup, which often provides inert atmosphere during the measurement, is less challenging for reduced air sensitivity and makes experiments with a high throughput of samples much easier. 
Furthermore, single crystals are often more sustainable and performing experiments after extended periods of storage or shipping is more likely possible than in case of their polycrystalline counterparts.  
\\
The first step to successful synthesize compounds containing reactive and volatile materials is finding a suitable container material.
Once this is done, we can take advantage of the many benefits of solution growth: reduced reaction temperatures, reduced vapor pressure, work with small amounts of material (compared to using mirror furnaces, Czochralski or Bridgeman methods), and "in-situ purification" of the starting materials often associated with crystallization from the melt. 
\\ 
In this paper, we will demonstrate the use of Li and Ca flux for the single crystal growth of materials containing light, volatile, and reactive materials. 
Given that both Li and Ca can act as excellent solvents for nitrogen, a special focus will be given to the growth of nitrides. 

\section{Experimental methods}

\begin{figure}
\center
\includegraphics[width=0.9\textwidth]{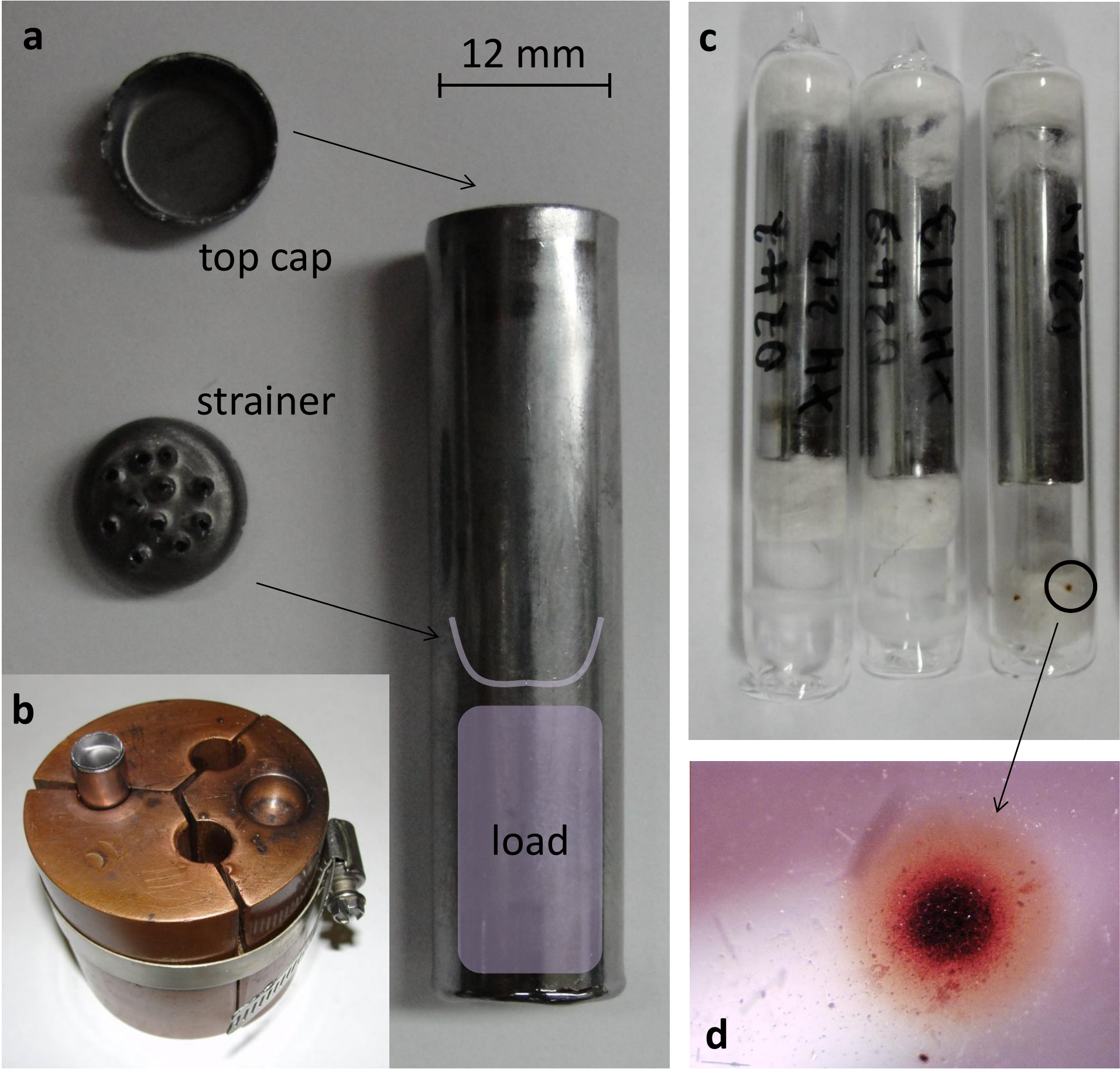}
\caption{a, Three-cap Ta crucible with top cap and strainer shown separately to left side. 
b, Massive copper crucible holder for arc welding up to three Ta crucibles in one run. 
c, Quartz ampules prevent the oxidization of the Ta crucible.
The right one shows black-reddish points due to attack. 
d, Magnified view of such a point which is likely caused by a reaction with Li or Ca which escaped through micro cracks along the welded seam.}
\label{cruc}
\end{figure}

Given the reactivity of Li and Ca with silica, we tested Ta and Nb as crucible materials and found both suitable for a variety of compounds and temperature profiles. 
Ta and Nb, in form of commercially available tubes and sheets, were machined to so called "three-cap crucibles"\,\cite{Canfield2001} which allow for easy decanting of excess flux (Fig.\ref{cruc}a). 
A massive copper crucible holder is used to mount the Ta crucibles and provide thermal inertia to keep the elements cool during welding (Fig.\ref{cruc}b).
One of the caps (bottom) is arc welded into one side of the tube (all arc welding is done under roughly 0.6 atmosphere Ar).
After loading the material, a second cap with holes drilled through it is placed above the solid material roughly in the middle of the tube. 
We found that no further fixation of this cap is necessary provided it fits tightly into the tube. 
The third cap (top) is arc welded into the top of the tube, sealing the crucible. 
Care must be taken to prevent a significant heating-up of any volatile materials during this last step.
Therefore, the arc welding was performed as quickly as possible and was completed in typically less than one minute. Optical inspection and weighing of the filled crucible before and after arc welding can reveal any signs of vapor transport out of the crucible.
Finally, the completed three-cap crucible is sealed in a silica ampule under roughly 1/3 atmosphere Ar to prevent oxidization of Ta or Nb (Fig.\ref{cruc}c).

Crystal growth attempts in such three-cap crucibles using Li and Ca flux were performed at temperatures of up to $1200^\circ$\,C.
After the samples were heated, the silica ampules occasionally showed black-reddish points on the level of the welded seam (Fig.\ref{cruc}d). 
These are likely caused by a reaction with Li or Ca which escaped through micro cracks along the welded seam that might open up only at elevated temperatures.
These are still relatively minor leaks and do not significantly hinder crystal growth.
This can be contrasted with an actual failure when a faulty sealing resulted in Li substantially attacking the silica ampule. 
The loss of the silica ampule's integrity lead to a complete disintegration of both Ta crucible and silica ampule and even a surrounding 50\,ml Al$_2$O$_3$ support crucible and an underlying hearth plate were damaged. 
In turn, this sensitivity to larger leaks in the Ta or Nb crucibles, as was demonstrated in this single, failed attempt, shows the general goodness of the sealing obtained in all the other, successful growth attempts.

X-ray diffraction (XRD) was performed using a Rigaku Miniflex diffractometer (wavelength: Cu-$K\alpha_{1,2}$).
Part of the measurements were performed under nitrogen atmosphere (see figure captions). 
Lattice parameters were refined by the LeBail method using GSAS\,\cite{Larson2000} and EXPGUI\,\cite{Toby2001}.
Instrument parameter files were determined from measurements on Si and Al$_2$O$_3$. 
XRD powder diffraction measurements on these standard materials have been performed regularly and allow to estimate the relative error for the given lattice parameters to be less than 0.002. 

In addition, X-ray diffraction was performed on some of the grown single crystals that show plate-like habit [LiIr, Ca$_2$N, Y$_2$Ni$_7$, and ({\bf Ca},Y)Ni$_3$ - see below]. For these measurements, the surface normal, which is parallel to the crystallographic $c$-direction, was oriented parallel to the scattering vector. In this way, only $0\,0\,l$ reflections are accessible. 
Laue-back-reflection patterns were taken with an MWL-110 camera manufactured by Multiwire Laboratories. 
Magnetization measurements were performed using a Quantum Design Magnetic Property Measurement System (MPMS).
Electron microscopy and energy dispersive X-Ray analysis were performed using a JOEL 59101LV system.  

\section{Lithium flux}
With a density of 0.53\,g/cm$^3$ lithium is the lightest solid element at ambient conditions. 
The low melting temperature of 180$^\circ$C in combination with a large separation from its boiling temperature ($1342^\circ$C) fulfill two key characteristics of a good flux.
However, the high reactivity and air-sensitivity of lithium requires particular attention. 
A moderately high vapor pressure ($10^{-2}$\,mbar at $520^\circ$C\,\cite{Honig1969}) is sufficient to cause a significant attack of silica glass via vapor transport of the Li to the silica.  
As shown above, sealed Ta crucibles are convenient up to temperatures of $1200^\circ$C.

Lithium has a low solubility for many transition metals in binary melts, e.g., well below 1\,at.\% for V, Cr, Mn, Fe, Co, Ni at $T = 1000^\circ$\,\cite{Lyublinski1995}.
However, the presence of a third element can increase the solubility significantly. This holds true for low level impurities \,\cite{Lyublinski1995} as well as larger amounts of a third element (see below). 
A rather good solubility is observed for the noble metals, e.g. for Pd\,\cite{Sangster1990b} and Pt\,\cite{Sangster1990}.
A brief review of the growth of intermetallic compounds from lithium flux is given in Ref.\,\cite{Kanatzidis2005}.

\subsection{Nitrides}
\begin{figure}
\center
\includegraphics[width=0.85\textwidth]{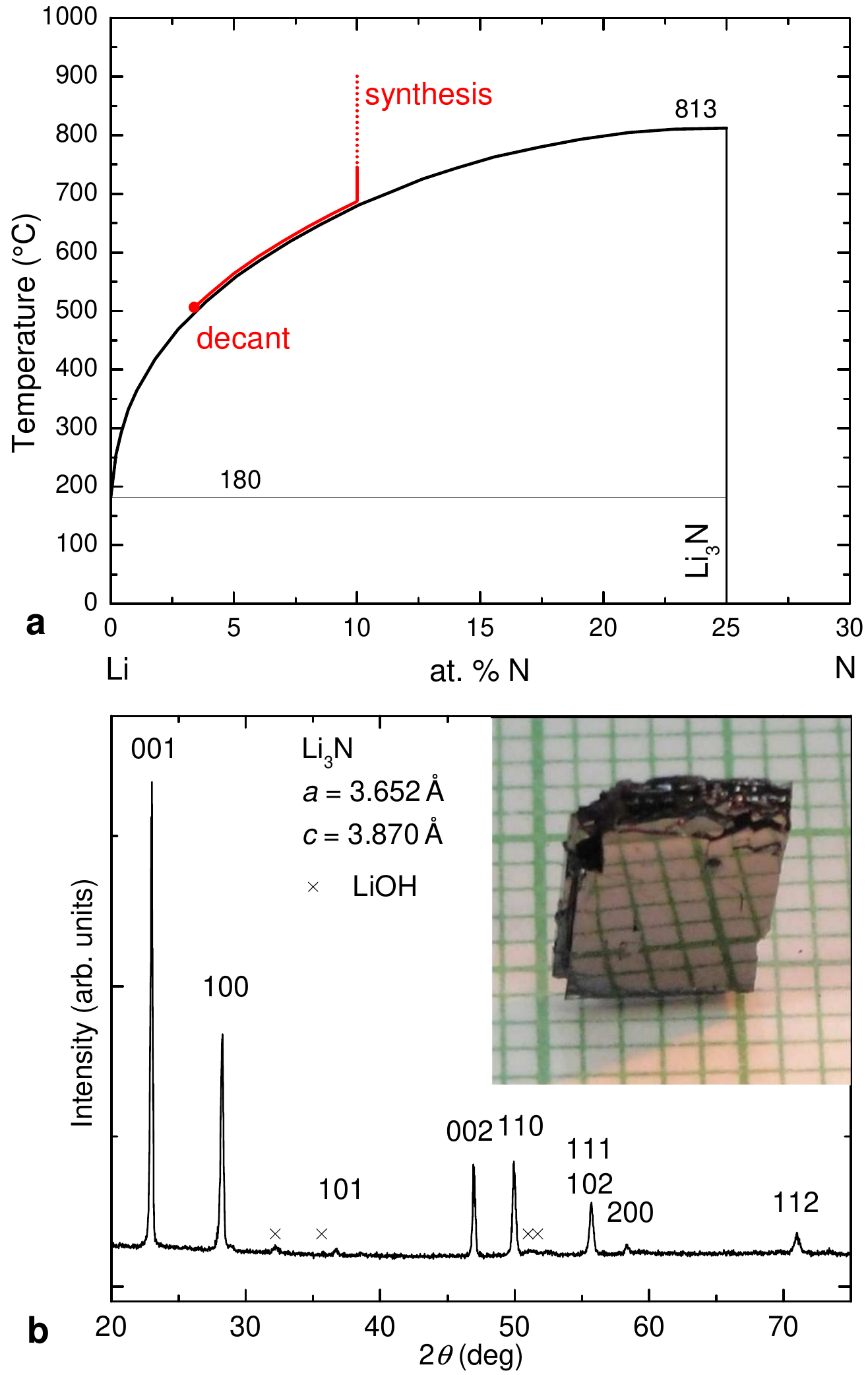}
\caption{a, Li-N phase diagram after Okamoto\,\cite{Okamoto1990LiN}. 
Typical starting composition, temperature profile, and composition of the liquid phase are shown schematically with the red line (the dotted line represents rapid cooling). 
b, X-Ray powder diffraction pattern measured on ground Li$_3$N single crystals under nitrogen atmosphere. 
A representative Li$_3$N single crystal is shown on a millimeter grid (roughly broken along the sides).
}
\label{phase_lin}
\end{figure}

The high solubility of nitrogen in liquid Li is demonstrated in the binary alloy phase diagram (Fig.\,\ref{phase_lin}a after Okamoto\,\cite{Okamoto1990LiN}).
It makes Li a promising flux for nitride single crystal growth at comparatively low temperatures of $800^\circ$C and below. Furthermore, other than Li$_3$N, there are no other Li-N binaries on the Li-rich side that could compete with the formation of other nitrides. 
  
\subsubsection{Li$_3$N}
The first step to grow nitrides out of Li-rich flux was to check the suitability of the Ta crucibles in the relevant temperature range and to verify the published binary alloy Li-N phase diagram in the framework of our experiment, in particular to make sure there is not excessive N$_2$ partial pressure over the melt. 
This was tested by the syntheses of Li$_3$N from a Li$_{90}$N$_{10}$ mixture put together out of Li lumps (Alfa, 99\,\%) and Li$_3$N powder (Alfa 99.4\,\%).
The Li-Li$_3$N mixture was heated from room temperature to $T = 900^\circ$C over 5\,h, cooled to T = 750$^\circ$C within 1.5\,h, slowly cooled to T = 500$^\circ$C over 50 to 100\,h, and finally decanted to separate the Li$_3$N crystals from the excess liquid. 
Plate-like single crystals with lateral dimensions limited by the crucible (12\,mm) and thickness of up to 2\,mm were obtained. A representative Li$_3$N single crystal is shown in Fig.\,\ref{phase_lin}b on a millimeter grid (roughly broken along the sides perpendicular to the $a$-$b$ plane).
XRD measurements on ground single crystals (Fig.\,\ref{phase_lin}b) revealed lattice parameters of $a = 3.652$\,\AA~and $c = 3.870$\,\AA.
This in good agreement with the reported values of $a = 3.648(1)$\,\AA~and $c = 3.875(1)$\,\AA\,\cite{Rabenau1976}.

For the Li$_{90}$N$_{10}$ initial melt composition there was no indication of any significant vapor pressure (as can be inferred from deformation of the Ta crucible).
We did find, though, that reducing the Li:N ratio to $\sim$4:1 (i.e. adding more nitrogen to the melt) and/or heating to temperatures above $1000^\circ$C resulted in a significant attack of the Ta crucibles which became very brittle even though no leaks occurred. 
In contrast, the Ta crucibles remained metallic, flexible and showed no signs of attack when using temperature profile and melt compositions as described above.

\subsubsection{Li$_2$(Li$_{1-x}$\textit{T}$_x$)N with $T$ = Mn, Fe, Co}

\begin{figure}
\center
\includegraphics[width=0.57\textwidth]{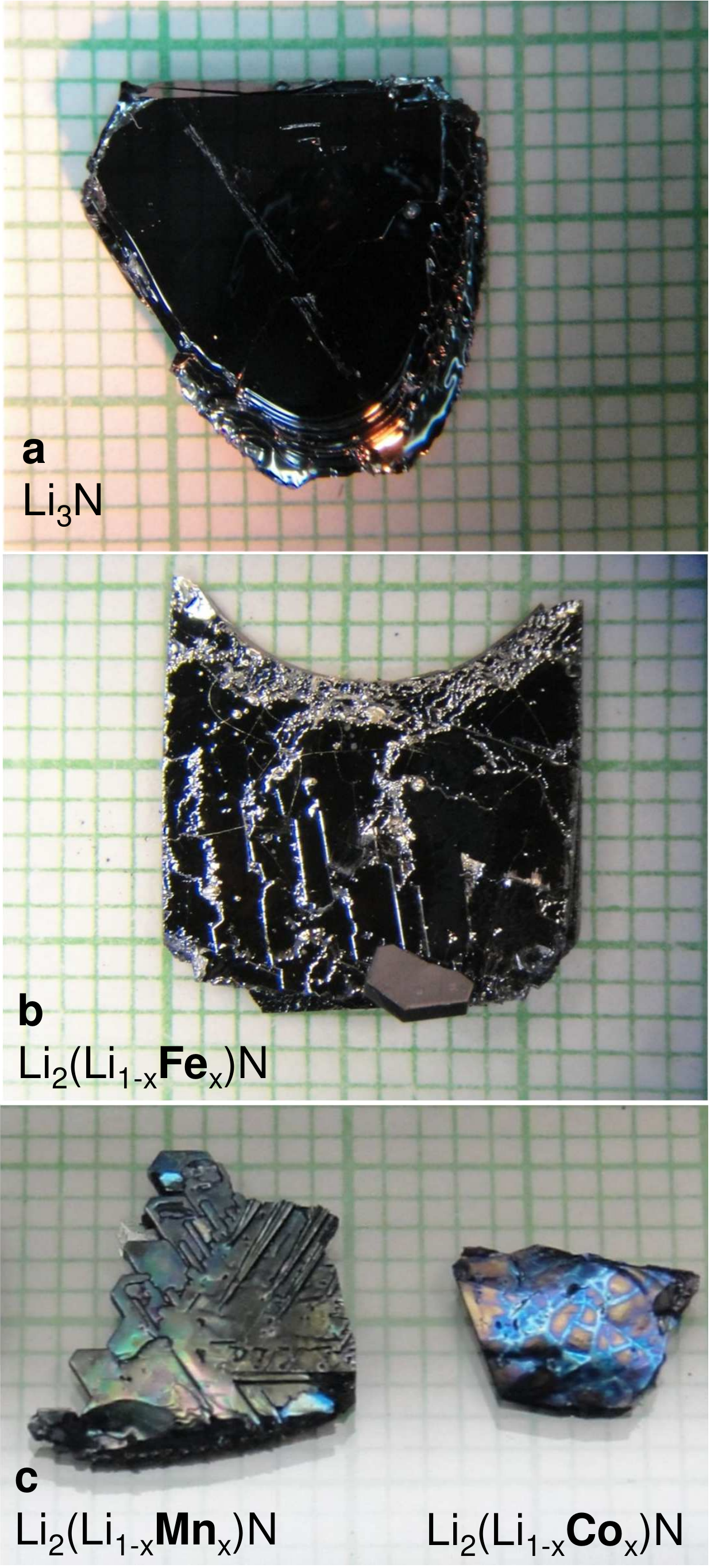}
\caption{
As-grown single crystals of nominally pure and doped Li$_3$N on a millimeter grid. 
a, Nominally pure Li$_3$N single crystal (actual composition being $\sim$\,Li$_{2.9999}$Fe$_{0.0001}$N - see text) obtained by slowly cooling a Li$_{90}$N$_{10}$ mixture from 750$^\circ$C over 67\,h to 550$^\circ$C. 
b, Li$_2$(Li$_{1-x}$Fe$_x$)N single crystal decanted at a lower temperature of $T = 500^\circ$C showing a crucible-limited size. The Li$_{89}$Fe$_1$N$_{10}$ mixture was slowly cooled from 750$^\circ$C over 63\,h to 500$^\circ$C resulting in $x \sim 0.1$.
Doping Fe has no significant effect on the crystal habit when compared to nominally undoped Li$_3$N.
c, In contrast, adding as little as 0.3\,at.\% Mn or Co to the melt changes the crystal habit significantly (the temperature profile used in these cases was identical to the one of nominally pure Li$_3$N shown above).}
\label{li3n_Mn-Co_crystal_v2}
\end{figure}

Transition metals ($T$) can be incorporated in Li$_3$N where they occupy only one of the two the Li sites (the 1$b$ Wyckoff site which is two fold coordinated by nitrogen)\cite{Sachsze1949}. 
However, it is surprising that we were able to grow single crystals containing significant amounts of transition metals (e.g. $x = 0.28$ for Fe) despite their reported negligible solubility in Li. 
The formation of $T$-N or $T$-N-Li complexes in the liquid state apparently allow much more transition metal to be incorporated into the Li-rich flux. 

Extensive work has been performed on the Li$_2$(Li$_{1-x}$Fe$_x$)N series which shows an extreme magnetic anisotropy and coercivity as well as exotic tunneling effects of the magnetization\,\cite{Jesche2014b}. 
An important aspect is the absence of ferromagnetic or superparamagnetic Fe particles as shown by magnetization measurements\,\cite{Jesche2014b}. This would be difficult to achieve using alternative growth procedures.
The starting materials were mixed in a molar ratio of Li:Fe:Li$_3$N =  6-$x_0$:$x_0$:1 with $x_0$ ranging from 0 to 0.5 (corresponding to a melt stoichiometry of Li$_{85}$Fe$_{5}$N$_{10}$ for the latter one).
The mixtures with a total mass of roughly 1.3\,g were packed in a three-cap Ta crucible, heated from room temperature to $T = 900^\circ$C over 5\,h, cooled to T = 750$^\circ$C within 1.5\,h, slowly cooled to 500$^\circ$C over 60 to 100\,h, and finally decanted to separate the single crystals from the excess flux. 
For slow cooling rates the sample size was limited by the crucible.
The crystal habit for $T$ = Fe is similar to undoped Li$_3$N single crystals for the whole concentration range ($x = 0.00013 - 0.28$). Notice that even nominally undoped Li$_3$N shows Fe concentrations of $x = 0.0001$ to 0.0003 that originate from Fe impurities in either the starting materials or leached from the Ta crucibles\,\cite{Jesche2014b}.
As demonstrated in Figs.\,\ref{li3n_Mn-Co_crystal_v2}a,b and observed in several additional growth procedures there is little change in the crystal habit as Fe is incorporated in the melt.

In contrast, the crystal habit for $T$ = Mn and Co differs significantly from undoped Li$_3$N (Fig.\,\ref{li3n_Mn-Co_crystal_v2}c). 
The growth parameters were the same as for the growth of undoped Li$_3$N single crystal shown in Fig.\,\ref{li3n_Mn-Co_crystal_v2}a. 
Doping levels of as little as $x_0 = 0.03$ Mn or Co, corresponding to 0.3\,at.\% in the melt, lead to significant changes in size, aspect ratio and formation of facets. 
For $T$ = Mn much clearer hexagonal, edge faceting is apparent.
For $T$ = Co, the single crystals were smaller and significantly thinner ($\sim 0.05$\,mm) when compared with undoped Li$_3$N or $T$ = Fe.
Magnetization measurements confirm the presence of Mn and Co. 

Attempts to grow Li$_2$(Li$_{1-x}$\textit{T}$_x$)N for $T$ = Sc, Ti, V, Cr failed. The product consisted of undoped Li$_3$N and, partially, the unreacted transition metals were still visible with the shape of the starting material preserved. 
To our knowledge, an oxidation state of predominantly +1 is not observed for Sc, Ti, V, or Cr in solids and this may be the origin for the failure of substituting Li by these elements. 
This can be viewed as being consistent with the idea of an oxidation state of +1 for $T$ = Mn, Fe, Co in these compounds which, although unusual, is reasonable for replacing Li$^{+1}$.
For $T =$ Cu and Ni this has been demonstrated by X-ray absorption spectroscopy\,\cite{Niewa2003}. 

A negative aspect of the surprising solubility of some transition metals in the Li$_{90}$N$_{10}$ melt is the possibility of also increasing the solubility of Ta or Nb in Li which is basically zero in the absence of nitrogen. 
This may be partially responsible for the change of the crucible materials as observed at higher temperatures.

\subsubsection{AlN}
Polycrystalline AlN was obtained as a result of an attempt to grow either AlN or AlLi$_3$N$_2$ from a Li-rich flux.
The formation of AlN from elemental Al dissolved in a Li$_{94}$N$_6$ melt is an important result since it proves that the Li$_3$N binary compound is not so stable that it prevents the growth of other (Li-free) nitrides.
It is reasonable to assume that this is not restricted to Al and that Li flux can be used as an exploratory tool for the growth of novel nitrides as well as for the growth of large single crystals of known compounds.

The starting materials were mixed in a molar ratios of Li:Al:Li$_3$N = 10:2:2 (corresponding to Li$_{71}$Al$_{14}$N$_{14}$).
The mixture with a total mass of roughly 1.5\,g was packed in a three-cap Ta crucible, heated from room temperature to $T = 900^\circ$C over 4\,h, slowly cooled to 400$^\circ$C over 44\,h, and finally decanted. 
Polycrystalline AlN was found at the bottom of the Ta crucible.
Specular reflecting facets of $\sim 50\,\mu$m along a side were observed among the polycrystalline material. These are likely attributed to small AlN single crystals suggesting that even more diluted melts, along with slower cooling might be needed for growth of larger AlN grains.

\begin{figure}
\center
\includegraphics[width=0.78\textwidth]{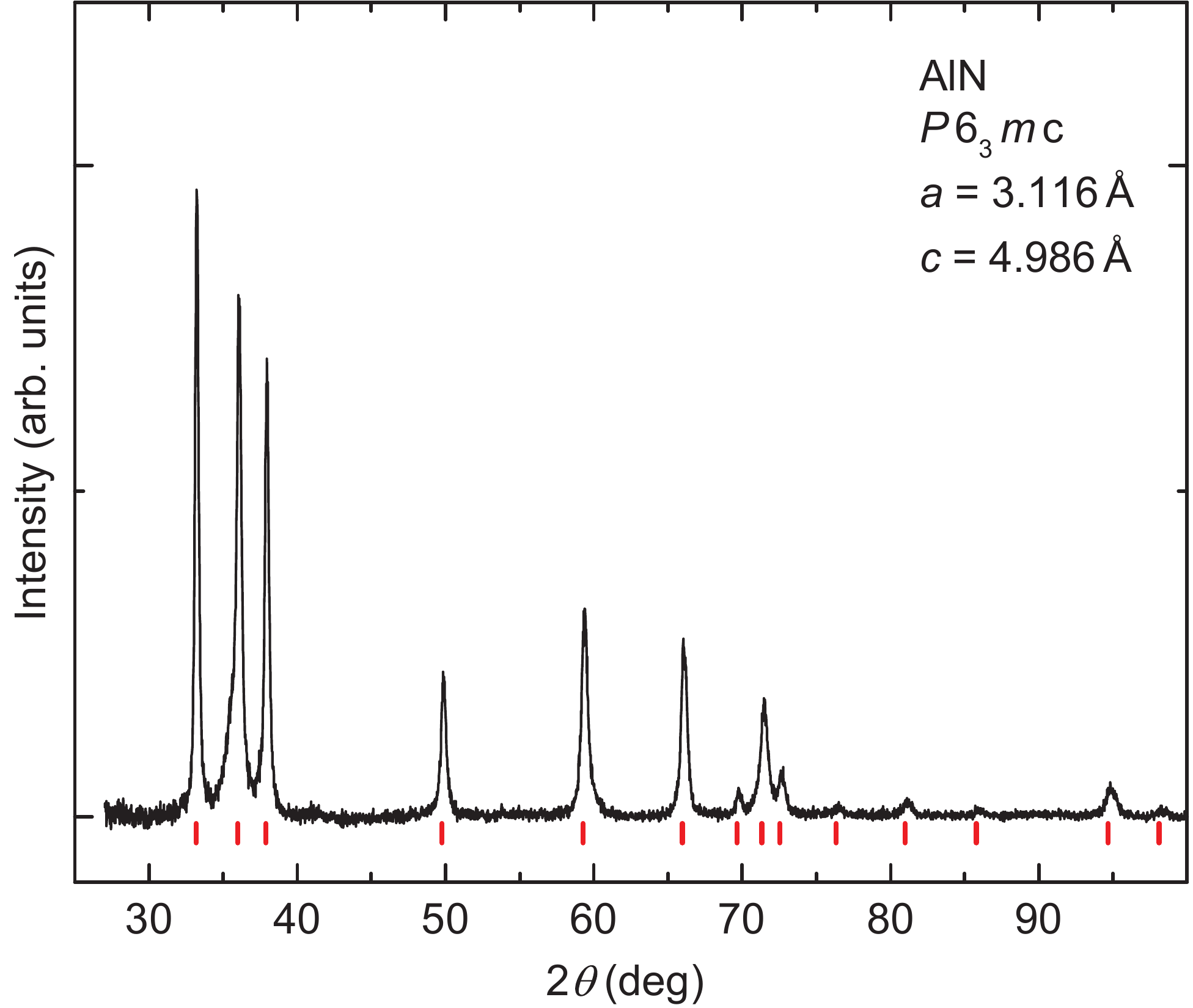}
\caption{X-Ray powder diffraction pattern collected under nitrogen atmosphere. The theoretical peak positions are marked by red bars.}
\label{diff-aln}
\end{figure}

Powder XRD measurements on AlN (Fig.\,\ref{diff-aln}) revealed lattice parameters of $a = 3.116$\,\AA~and $c = 4.986$\,\AA.
This in good agreement with the reported values of $a = 3.1115$\,\AA~and $c = 4.9798$\,\AA\,\cite{Yim1974}.

\subsubsection{LiCaN}
\begin{figure}
\center
\includegraphics[width=0.65\textwidth]{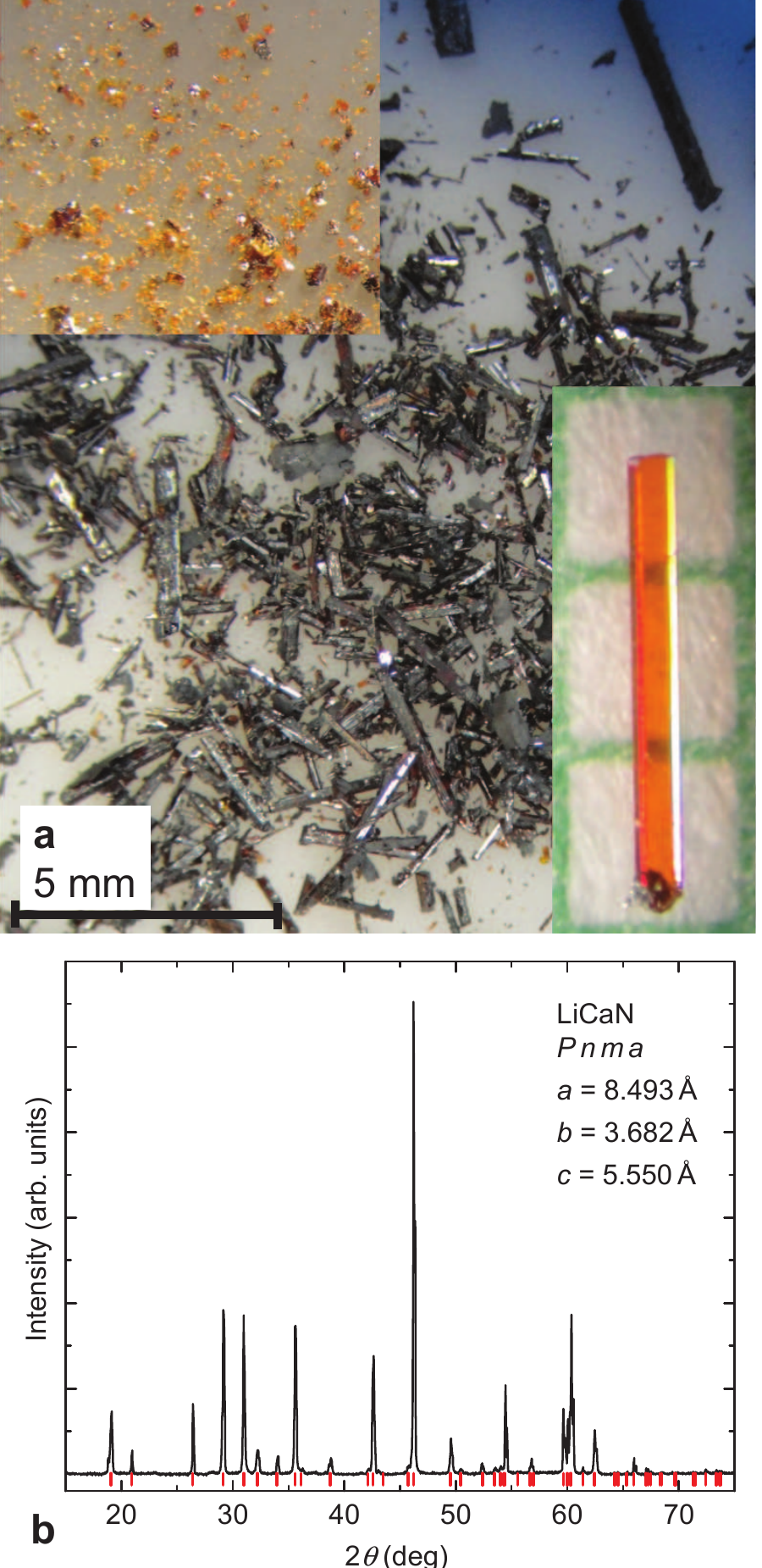}
\caption{a, Single crystals of LiCaN grown from a Li$_{82}$Ca$_{9}$N$_{9}$ melt, partially coated with grayish flux remnants. The orange color becomes apparent after breaking or grinding the samples (upper left part). An as-grown LiCaN single crystal obtained as a by-product from a Li$_{90}$N$_{10}$ melt is shown on a millimeter grid (lower right part, Ca was introduced as an impurity - see text).
b, X-Ray powder diffraction pattern collected under nitrogen atmosphere. The theoretical peak positions are marked by red bars.}
\label{LiCaN}
\end{figure}  

Initially, we observed LiCaN as a by-product of the growth of Li$_3$N where Ca was introduced as an impurity phase of the starting materials (Alfa, Li lumps 99\,\%, Li$_3$N powder 99.4\,\%). The amount of LiCaN can be estimated to less than 1\,\% of the growth product, consistent with the purity of the starting materials.
Later, we added Ca intentionally to the melt in order to grow larger quantities of this material. 
The starting materials were mixed in a molar ratio of Li:Ca:Li$_3$N = 6:1:1 (corresponding to Li$_{82}$Ca$_{9}$N$_{9}$).
The mixture with a total mass of roughly 1.4\,g was packed in a three-cap Ta crucible, heated from room temperature to $T = 900^\circ$C over 4.5\,h, cooled to T = 750$^\circ$C within 1.5\,h, slowly cooled to 350$^\circ$C over 90\,h, and finally decanted to separate the LiCaN crystals from the remaining liquid. 

Rod-shaped single crystals with lengths of up to 5 mm and cross-sectional dimensions of up to 1\,mm were obtained (Fig.\ref{LiCaN}a). 
The orange color is clearly visible in crushed single crystals (upper, left inset in Fig.\ref{LiCaN}a) whereas the color of the as grown samples often appears more faint due to flux remnants on the sample surface. 
A LiCaN single grown as a by-product from a nominally Ca-free Li$_{90}$N$_{10}$ solution, i.e. a very dilute Li$_{90-\delta}$Ca$_\delta$N$_{10}$ solution, is shown in the lower, right inset of Fig.\ref{LiCaN}a. The surface is cleaner and the orange color more apparent when compared to the samples grown from a more Ca-rich flux. 

Powder XRD measurements on ground LiCaN single crystals (Fig.\,\ref{LiCaN}b) revealed lattice parameters of $a = 8.493$\,\AA, $b = 3.682$\,\AA~and $c = 5.550$\,\AA.
The reported values of $a = 8.471(3)$\,\AA, $b = 3.676(2)$\,\AA~and $c = 5.537(3)$\,\AA~are somewhat smaller\,\cite{Cordier1989}. These values were obtained by means of single crystal diffraction on samples grown from a Li$_6$Ca mixture under nitrogen atmosphere. Furthermore, a plate-like habit is reported\,\cite{Cordier1989} whereas we observe a pronounced rod-like habit. The cause of the observed differences is unclear so far.

\subsection{Non-nitrides}
In addition to its ability to bring nitrogen into a useful solution, Li can act as a solvent for other elements as well. 
In this section we present proof of principle growths for a representative selection of other binary melts.

\subsubsection{Li$_2$C$_2$}
\begin{figure}
\center
\includegraphics[width=0.9\textwidth]{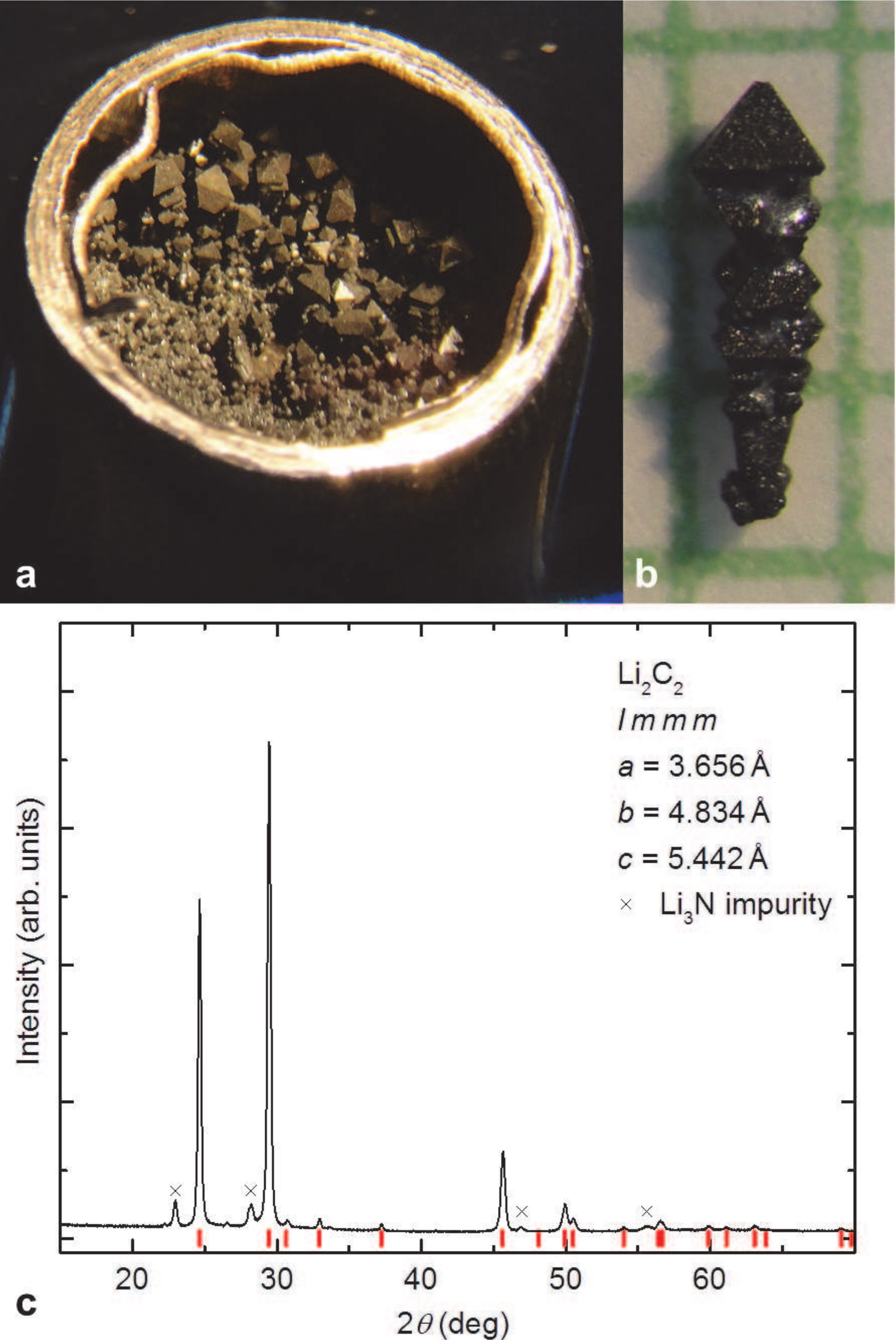}
\caption{a, Li$_2$C$_2$ single crystals in the opened bottom part of a three-cap Ta crucible. b, An isolated crystal is shown on a millimeter grid.
c, X-Ray powder diffraction pattern collected under nitrogen atmosphere. 
The theoretical peak positions are marked by red bars.
The Li$_3$N impurity likely originates from Li-rich flux remnants which were attached to the Li$_2$C$_2$ single crystals (see silver-grey droplets in b) that decay to Li$_3$N under the nitrogen atmosphere surrounding the X-Ray system.}
\label{LiC_1}
\end{figure} 

In order to evaluate the feasibility of a Li-rich flux for the growth of carbides we grew Li$_{2}$C$_{2}$ out of a binary melt.
As described above for the case of Li$_3$N, the following key questions had to be answered: is the Ta-crucible attacked, is the solubility sufficient, does the crystallization take place in a controlled fashion? 
Informally the question is, "Can we get carbon IN and OUT of solution?" 
The growth of Li$_2$C$_2$ single crystals of roughly one millimeter along a side provides a positive answer to all of these questions.
 
The starting materials were mixed in a molar ratio of Li:C = 9:1.
The mixture with a total mass of roughly 0.8\,g was packed in a three-cap Ta crucible, heated from room temperature to $T = 1000^\circ$C over 5\,h, held for 1\,h, slowly cooled to 400$^\circ$C over 60\,h, and finally decanted to separate the Li$_2$C$_2$ crystals from the excess liquid.  
The product is air-sensitive in powder form and decays on a timescale of one day.
XRD measurements on ground Li$_2$C$_2$ single crystals (Fig.\,\ref{LiC_1}c) revealed lattice parameters of $a = 3.656$\,\AA, $b = 4.834$\,\AA~and $c = 5.442$\,\AA.
This in good agreement with the reported values of $a = 3.655$\,\AA, $b = 4.830$\,\AA, $c = 5.440$\,\AA\,\cite{Juza1965} and $a = 3.6520(1)$\,\AA, $b = 4.8312(2)$\,\AA, $c = 5.4344(1)$\,\AA\,\cite{Ruschewitz1999}.

\subsubsection{Li$_3$Al$_2$}
\begin{figure}
\center
\includegraphics[width=0.9\textwidth]{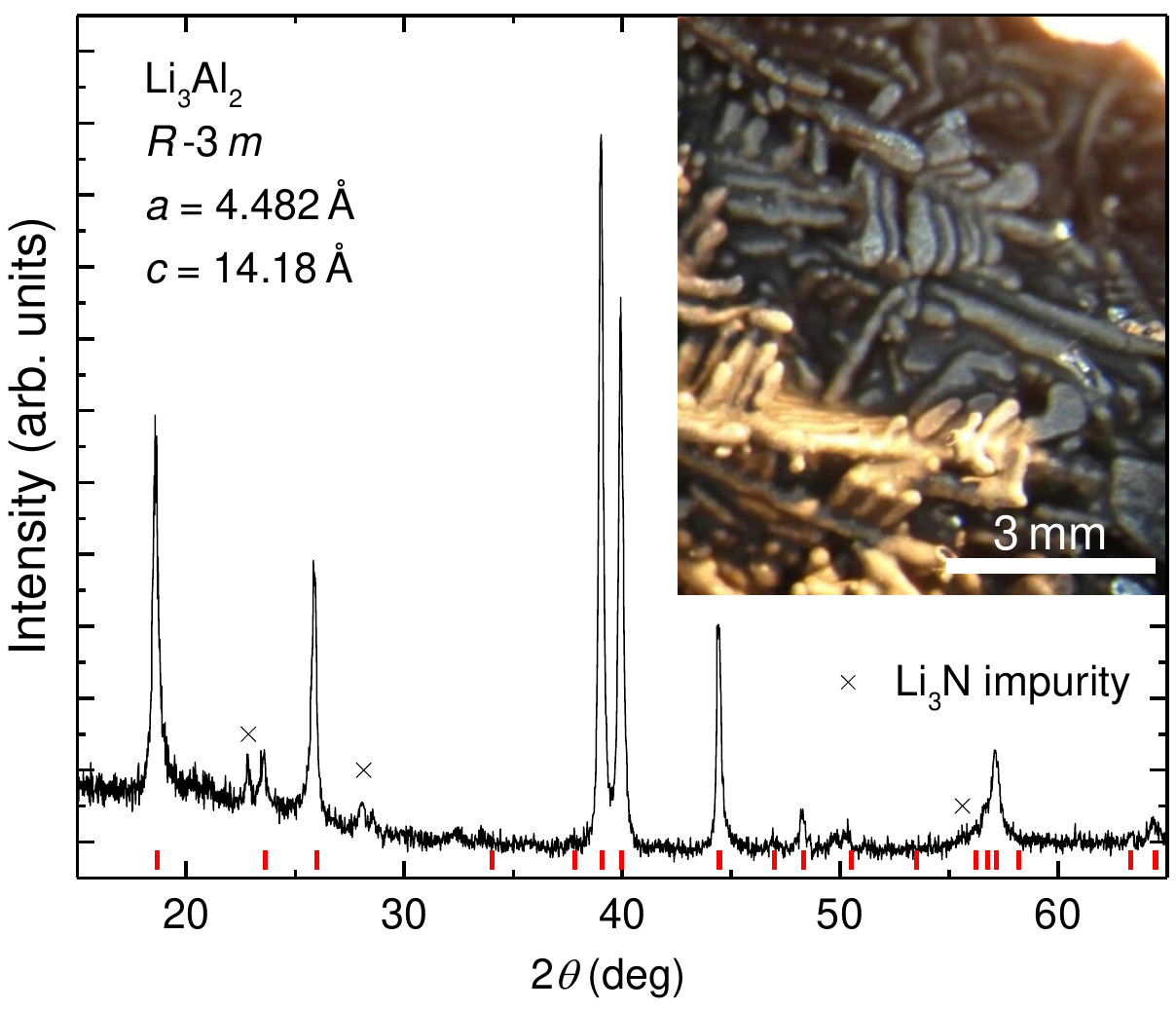}
\caption{
X-Ray powder diffraction pattern collected under nitrogen atmosphere. 
The theoretical peak positions for Li$_3$Al$_2$ are marked by red bars.
A small amount of a Li$_3$N impurity phase likely stems from Li-rich flux remnants which decay to Li$_3$N under nitrogen atmosphere.
The inset shows Li$_3$Al$_2$ dendrites grown from Li-rich flux. 
}
\label{li3al2}
\end{figure}  

The Li-Al binary phase diagram contains two Li-rich line compounds: Li$_3$Al$_2$ and Li$_9$Al$_4$\,\cite{Hallstedt2007}. In order to test the viability of the Li-Al melt we grew Li$_3$Al$_2$. 
An attempt to grow single crystals from a Li$_{80}$Al$_{20}$ mixture yielded dendritic structures at the bottom of the Ta crucible (Fig.\,\ref{li3al2}).
The mixture with a total mass of roughly 1.1\,g was packed in a three-cap Ta crucible, heated from room temperature to $T = 600^\circ$C over 3\,h, held for 2\,h, slowly cooled to 360$^\circ$C over 60\,h, and finally decanted. 
 
\begin{figure}
\center
\includegraphics[width=0.8\textwidth]{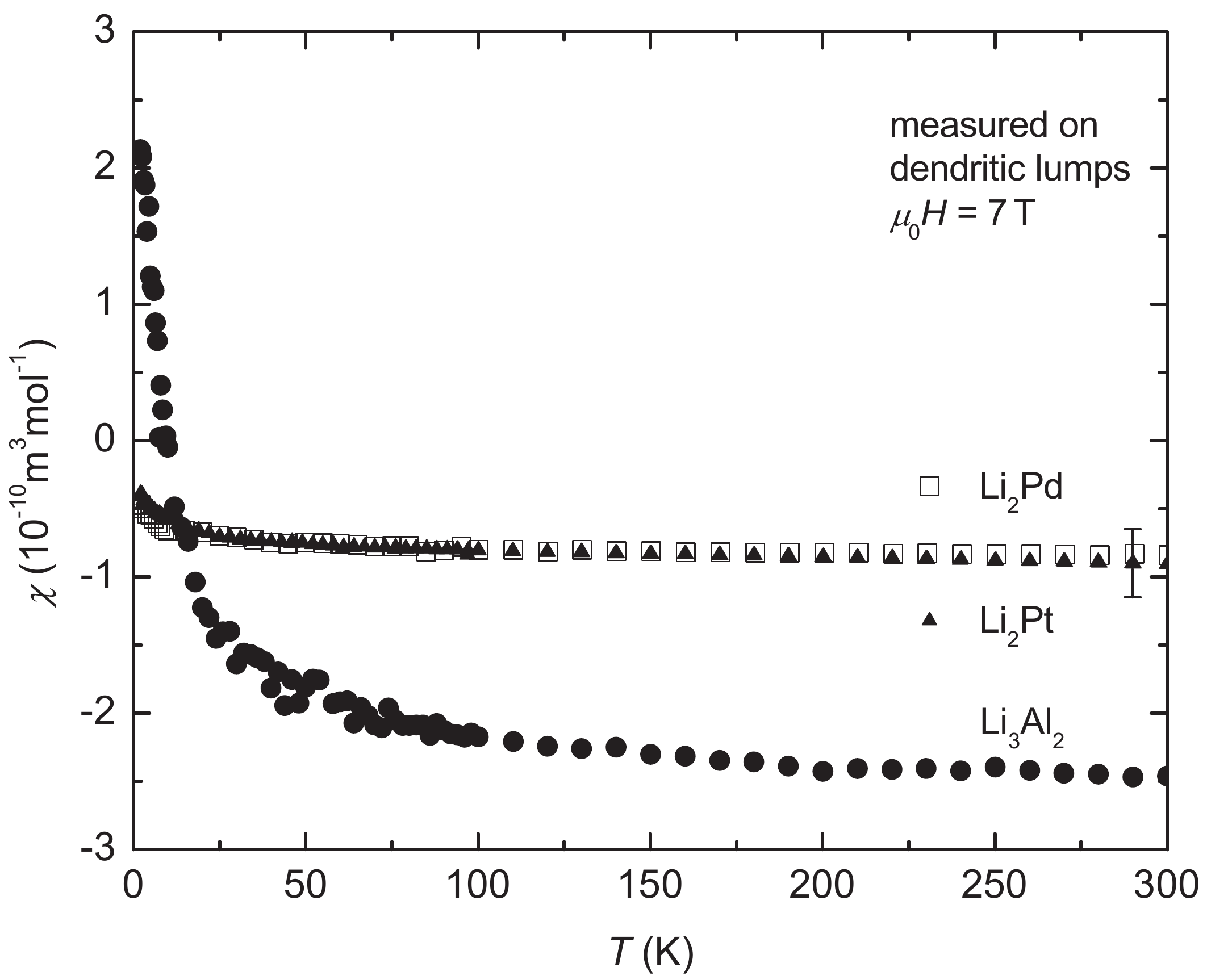}
\caption{Temperature dependent magnetic susceptibility, $\chi = M/H$, of Li$_3$Al$_2$, Li$_2$Pd, and Li$_2$Pt. 
The compounds are diamagnetic with a roughly temperature-independent susceptibility for $T = 50$ to 300\,K. 
The upturn observed upon cooling below $\sim50$\,K, more pronounced for Li$_3$Al$_2$, is attributed to a small amount of a local moment bearing impurity.
The estimated error is shown, for example, for Li$_2$Pt at $T = 290$\,K).
}
\label{mag-dendrite}
\end{figure}

XRD measurements on ground Li$_3$Al$_2$ dendrites (Fig.\,\ref{li3al2}) revealed lattice parameters of $a = 4.482$\,\AA~and $c = 14.18$\,\AA.
This is somewhat smaller than the reported values of $a = 4.508(7)$\,\AA~and $c = 14.259$\,\AA\,\cite{Tebbe1973}.
Magnetization measurements on a cluster of Li$_3$Al$_2$ dendrites revealed a roughly temperature-independent diamagnetic susceptibility of $-2.4(2)\cdot10^{-10}$\,m$^3$\,mol$^{-1}$ for $T = 50$ to 300\,K (Fig.\,\ref{mag-dendrite}, measured in an applied magnetic field of $\mu_0H = 7$\,T). An increase of the susceptibility upon cooling below $\sim 50$\,K (Curie tail) can be attributed to a small amount of local moment bearing magnetic impurities.
No indications for phase transitions were observed. 
In particular, additional measurements performed in small applied fields revealed no signs of superconductivity for $T_{\rm c} > 2$\,K.

\subsubsection{Li$_2$Pd and Li$_2$Pt}
\begin{figure}
\center
\includegraphics[width=0.9\textwidth]{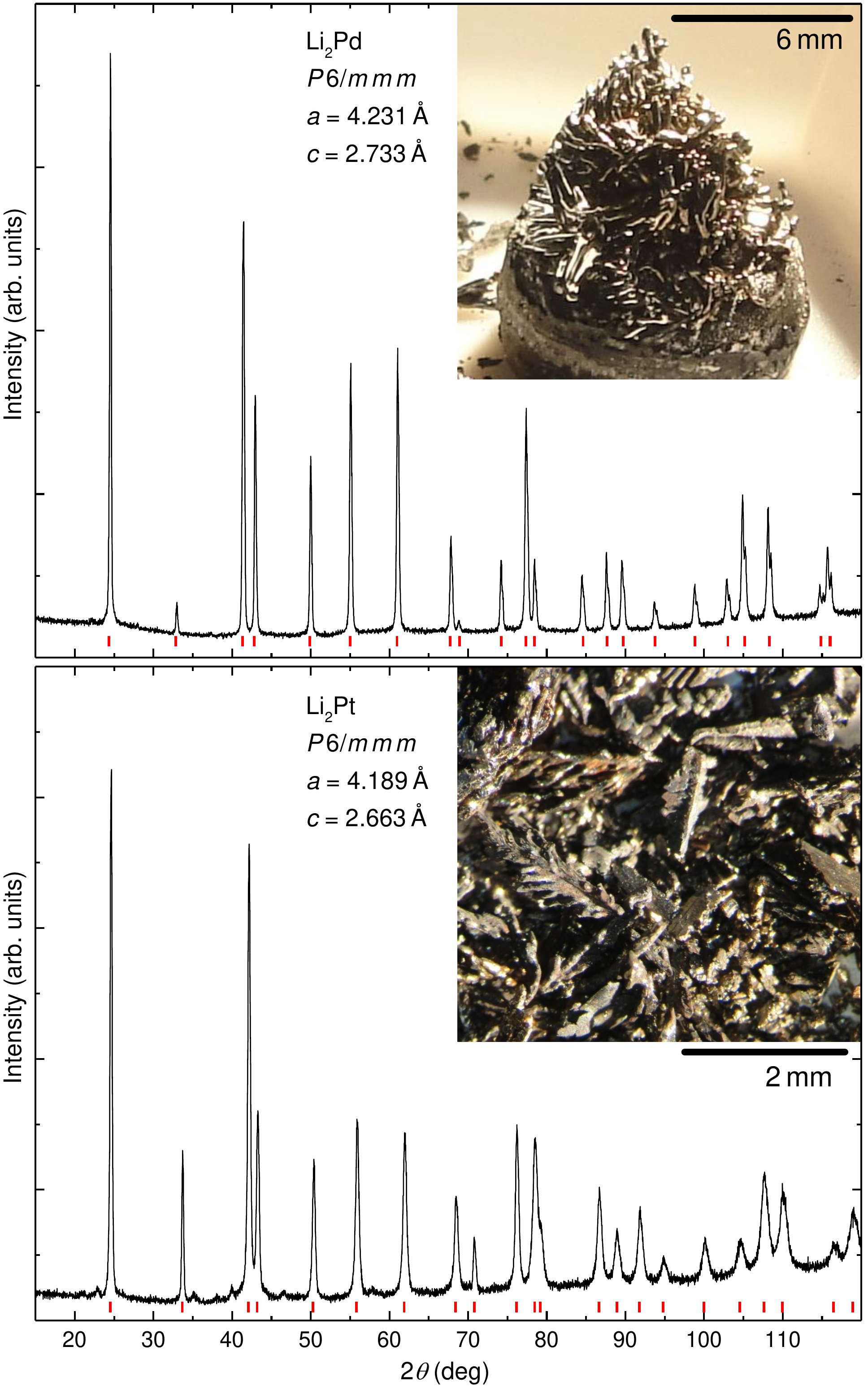}
\caption{
X-Ray powder diffraction pattern of Li$_2$Pd and Li$_2$Pt collected under nitrogen atmosphere. 
The theoretical peak positions are marked by red bars.
The insets show the pronounced dendritic structures as observed for both compounds.}
\label{dendritic}
\end{figure} 

In order to test the solubility of Pd and Pt in Li as well as our ability to contain Li-Pd and Li-Pt melts we chose Li$_2$Pd and Li$_2$Pt as target materials. They required moderate Pd and Pt concentrations as well as temperatures up to $\sim1000^\circ$C.
Our attempts to grow Li$_2$Pd and Li$_2$Pt single crystals from a Li-rich flux yielded phase clean material showing pronounced dendritic structures (Fig.\,\ref{dendritic}).

The starting materials for Li-Pd were mixed in a molar ratio of Li:Pd = 70:30 and Li:Pd = 73:27.
The mixtures with a total mass of roughly 4\,g were packed in three-cap Ta crucibles, heated from room temperature to $T = 1000^\circ$C over 5\,h, held for 0.5\,h, cooled to 610$^\circ$C over 4\,h, slowly cooled to 410$^\circ$C over 40\,h, and finally decanted. 
The starting materials for Li-Pt were mixed in a molar ratio of Li:Pt = 77:23 and Li:Pt = 80:20.
The mixtures with a total mass of roughly 4\,g were packed in three-cap Ta crucibles, heated from room temperature to $T = 1000^\circ$C over 5\,h, held for 2\,h, cooled to 800$^\circ$C over 2\,h, slowly cooled to 350$^\circ$C over 55\,h, and finally decanted. 
Slower cooling or use of powdered Pd and Pt or pre-arc-melted lumps of Pd and Pt had no significant effect on the growth result. 

XRD measurements on as-grown dendrites of Li$_2$Pd (Fig.\,\ref{dendritic}) revealed lattice parameters of $a = 4.231$\,\AA,~and $c = 2.733$\,\AA. 
Rather surprisingly, the measured intensities show no signs of a preferred orientation.
A measurement on ground dendrites revealed the same result. 
The obtained lattice parameters are in good agreement with the reported values of $a = 4.2267$\,\AA~and $c = 2.7319$\,\AA\,\cite{vanVucht1976}.
XRD measurements on ground dendrites of Li$_2$Pt (Fig.\,\ref{dendritic}) revealed lattice parameters of $a = 4.189$\,\AA,~and $c = 2.663$\,\AA. 
This is significantly smaller than the values obtained for Li$_2$Pd and in good agreement with the literature data of $a = 4.186$\,\AA,~and $c = 2.661$\,\AA\,\cite{Bronger1975}. 

Recent band-structure calculations propose that the Pt atoms in Li$_2$Pt exist as partially negative anions\,\cite{Lee2008} which is unusual for transition metals. 
However, this compound is, similar to isoelectronic Li$_2$Pd, rarely investigated experimentally.
Magnetization measurements performed on dendritic lumps revealed a roughly temperature-independent diamagnetic susceptibility of $-0.9(2)\cdot10^{-10}$\,m$^3$ mol$^{-1}$ for both Li$_2$Pd and Li$_2$Pt (Fig.\,\ref{mag-dendrite}). 
A small increase of the susceptibility upon cooling below $\sim 50$\,K (Curie tail) can be attributed to a small amount of local-moment bearing magnetic impurities.
No indications for phase transitions were observed.
In particular, additional measurements performed in small applied fields revealed no signs of superconductivity down to 2\,K.

\subsubsection{LiRh and LiIr}
\begin{figure}
\center
\includegraphics[width=0.9\textwidth]{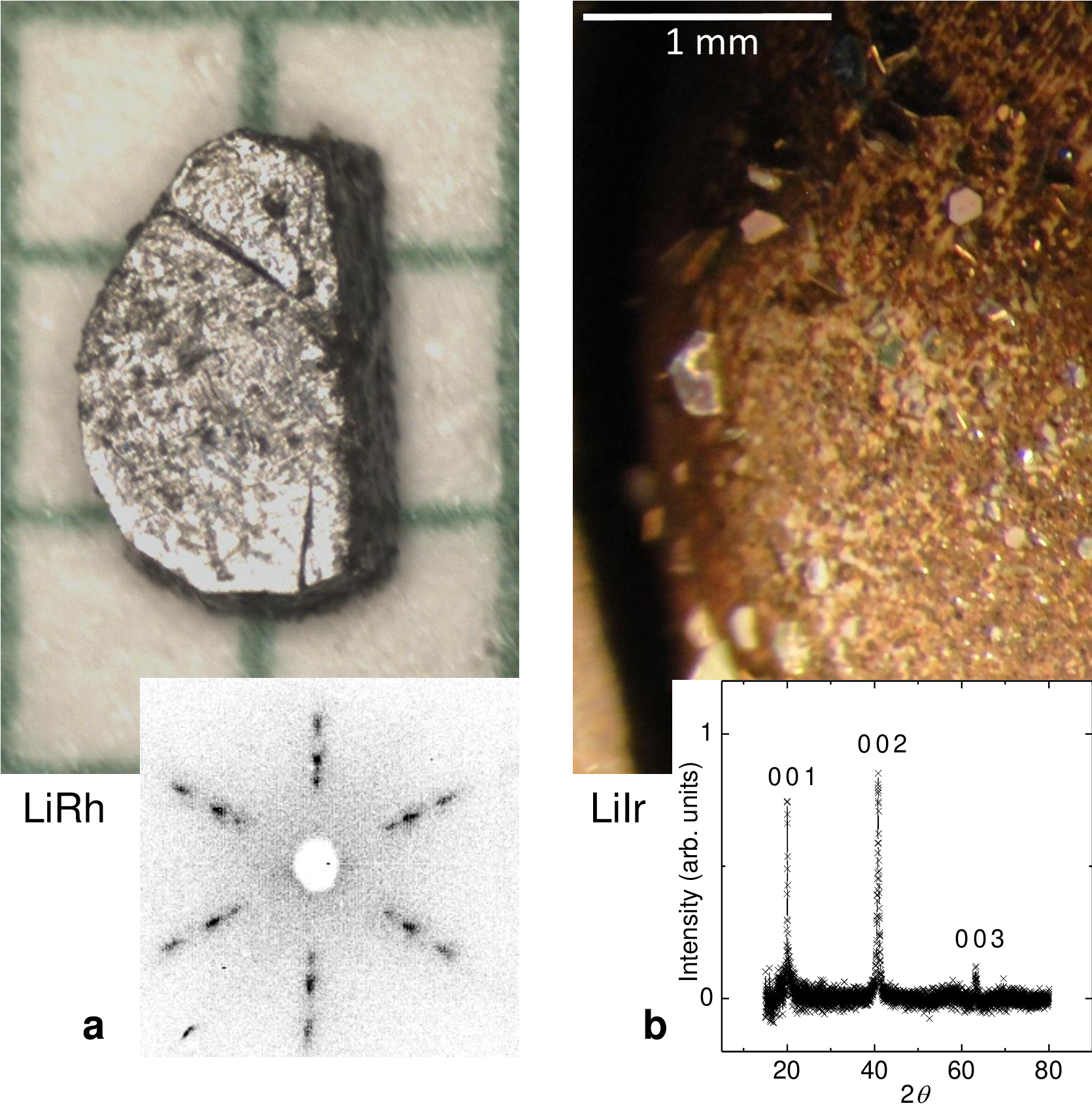}
\caption{a, Single crystal of LiRh on a millimeter grid and corresponding Laue-back-reflection pattern. b, LiIr single crystals attached to the bottom of a Ta crucible. The X-Ray diffraction pattern was recorded (in air) on a single crystal with the surface normal of the plate-like sample parallel to the scattering vector.}
\label{LiRh_LiIr2}
\end{figure} 

The published binary alloy phase diagrams for Li-Rh and Li-Ir indicate an uncertainty of the liquidus temperatures ('The assessed Li-Rh phase diagram is highly speculative'\,\cite{Massalski1990}). 
In order to get a rough sense on how well Rh and Ir dissolve in Li we performed attempts to grow LiRh and LiIr from a Li-rich flux. 
The starting materials for Li-Rh and Li-Ir were mixed in a molar ratio of Li:Rh/Ir = 95:5. 
Rh and Ir were arc melted prior to the growth in order to reduce the amount of hydrogen and other volatile impurities and provide recognizable lumps in the case of poor or no solubility.
The mixtures with a total mass of roughly 1.5\,g were packed in three-cap Ta crucibles, heated from room temperature to $T = 1000^\circ$C over 5\,h, held at $1000^\circ$C for 1\,h, cooled to 900$^\circ$C over 2\,h, slowly cooled to 250$^\circ$C over 90\,h, and finally decanted to separate the single crystals from the excess liquid. We succeeded in growing plate-like single crystals of LiRh with lateral dimensions of up to 2\,mm and thickness of $\sim0.5$\,mm using a Li-rich flux (Fig.\,\ref{LiRh_LiIr2}a).
The analogous attempt to grow LiIr yielded smaller single crystals up to 0.25\,mm along a side with a thickness of $\sim 20\mu$m (Fig.\,\ref{LiRh_LiIr2}b).
The significantly smaller size of the LiIr single crystals indicates differences in the nucleation and growth process when compared to LiRh. 
A possible reason is a higher liquidus temperature for Li$_{95}$Ir$_{5}$ in close vicinity to the maximum furnace temperature of $1000^\circ$C. This scenario would be consistent with the melting temperature of LiIr being roughly $200^\circ$C higher than the one of LiRh\,\cite{Massalski1990}.
Another very different possibility is a eutectic point in the vicinity of $\sim$Li$_{96}$Ir$_{4}$ (which is shifted to a higher Li concentration for LiRh).

The crystal structure of LiRh was studied by means of X-ray and neutron diffraction (space group $P\,\bar{6}$ originally reported\,\cite{Sidhu1965}, later revised to $P\,\bar{6}\,m\,2$\,\cite{Cenzual1991}). 
The synthesis and crystal structure of LiIr was reported in 1976\,\cite{Donkersloot1976} (isostructural to LiRh). 
Furthermore, both the Li-Rh-H\,\cite{Magee1964} and the Li-Ir-H\,\cite{Varma1978} system were investigated. 
Besides these early studies we are not aware of any report on physical properties of these rather simple binaries.
The sixfold crystallographic $c$-axis is oriented perpendicular to the larger surface of the plates as confirmed by Laue-back-reflection for LiRh and for LiIr by X-ray diffraction in a (symmetric) $\theta$-2$\theta$ geometry on oriented plates (the scattering vector is parallel to surface normal, the sample size of the LiIr single crystals is not sufficient for the available Laue-back-reflection setup).
Both, the LiRh and the LiIr single crystals were found to be extremely malleable. 
Accordingly, we did not succeed in grinding the material to a powder suitable for X-ray diffraction.  
However, a powder diffraction pattern of LiIr could be obtained from polycrystalline material which got stuck at the strainer of the used Ta crucible (not shown). The determined lattice parameters of $a = 2.650$\,\AA~and $c = 4.399$\,\AA~are in good agreement with the reported values [$a = 2.650(1)$\,\AA~and $c = 4.398(2)$\,\AA\,\cite{Donkersloot1976}].

\begin{figure}
\center
\includegraphics[width=\textwidth]{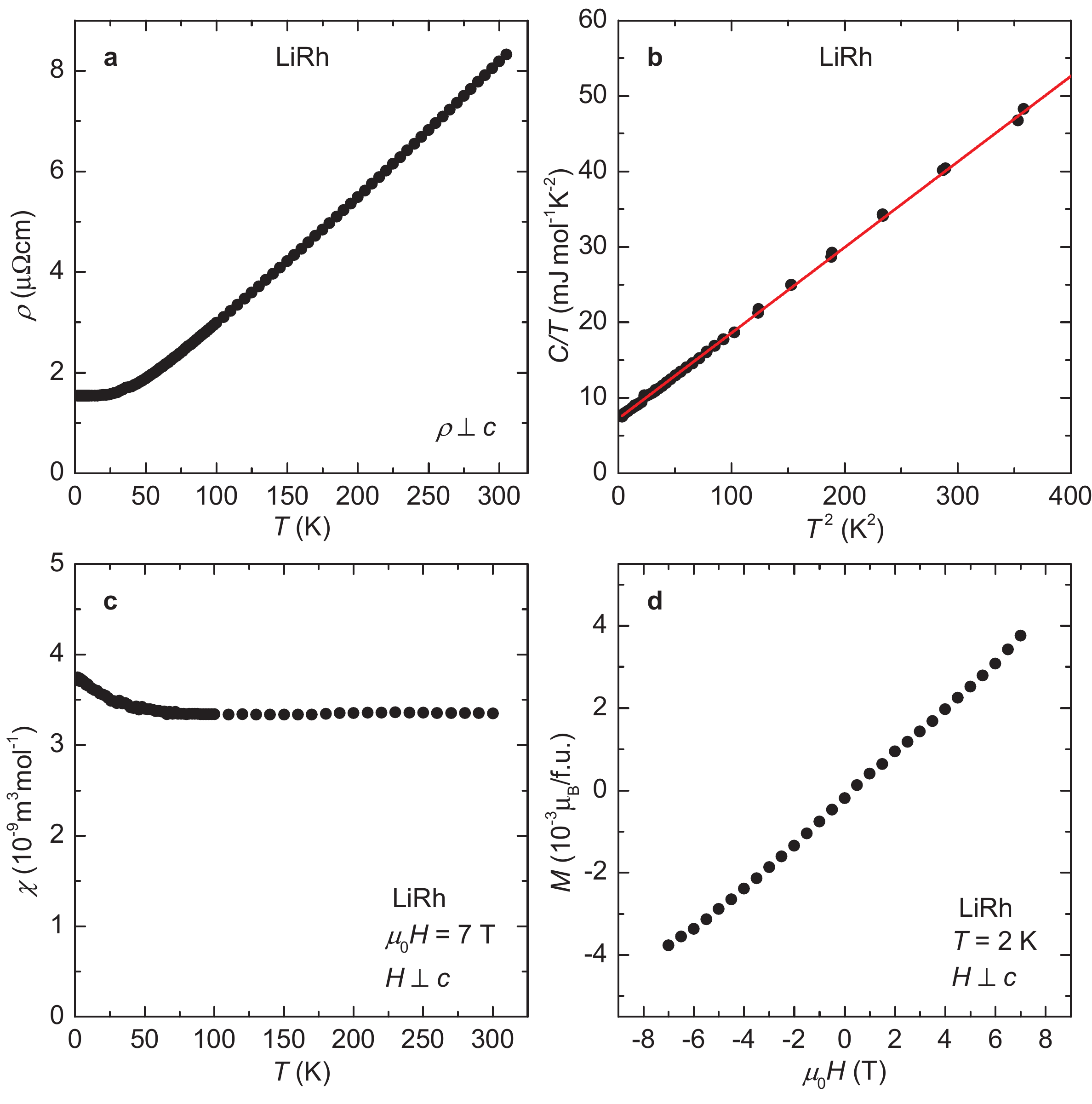}
\caption{Basic properties of LiRh. Temperature-dependent electrical resistivity (a), specific heat (b), and magnetization (c) as well as isothermal magnetization (d) indicate a magnetically non-ordered, metallic state for $T = 2$ to 300\,K.}
\label{lirh_phys}
\end{figure}

A basic characterization of LiRh by means of temperature-dependent electrical resistivity, magnetization and specific heat is presented in Fig.\,\ref{lirh_phys}. 
No indications for phase transitions were observed for $T = 2$ to 300\,K. 
The electrical resistivity shows the temperature-dependence of a normal metal with a residual resistivity ratio of RRR = 5 and a residual resistivity of $\rho_0 = 1.5\,\mu\Omega$\,cm (Fig.\,\ref{lirh_phys}a).
The low-temperature specific heat reveals a slightly enhanced Sommerfeld coefficient of $\gamma = 7.2$mJ\,mol$^{-1}$\,K$^{-2}$ (Fig.\,\ref{lirh_phys}b). The Debye temperature is estimated to $\Theta_{\rm D} = 182$\,K (low temperature limit). This rather small value is in accordance with the observation of small Debye temperatures in soft materials whereas hard materials show higher values (see e.g.\,\cite{Simon2013}).  
Temperature-dependent and isothermal magnetization show the typical behavior of a paramagnetic metal (Figs.\,\ref{lirh_phys}c,d).

\newpage
\section{Calcium flux}
With a density of $\rho = 1.55$\,g/cm$^3$ calcium is the lightest of the alkaline earth elements. 
It is highly reactive and air and moisture sensitive. 
This, as well as the moderately high melting temperature of $842^\circ$C and vapor pressure of 2\,mbar at $T = 842^\circ$C\,\cite{Honig1969} makes crystal growth from calcium flux a challenging task.
Promising, on the other hand, is the excellent solubility for the light elements N, C, and Al and further for the transition metals Ni, Cu, Pd, Ag, Pt, and Au in Ca (e.g.\,\cite{asmphase}).  

In section 4.1 we present results on the single crystal growth of binary calcium nitrides from Ca-rich melts. 
Several attempts to grow other nitrides from Ca-N-$X$ ternary melts failed, presumably due to the relatively high stability of binary Ca$_2$N. 
A brief report on these attempts, including CaNiN, CaMg$_2$N$_{2}$, Ca$_{3}$AuN, Ag$_{16}$Ca$_6$N, and Ag$_8$Ca$_{19}$N$_7$ is presented in section 5.
Section 4.2 summarizes our uses of Ca to bring Ni into solution and grow Ca-Ni binary compounds as well as representative $R$-Ni compounds ($R$ = Y, La, Yb).
As such, we believe that, in addition to its use as a solvent for light and reactive elements, Ca-rich flux is a suitable exploratory tool for the synthesis of rare earth bearing compounds at comparatively low temperatures.  

\begin{figure}
\center
\includegraphics[width=0.8\textwidth]{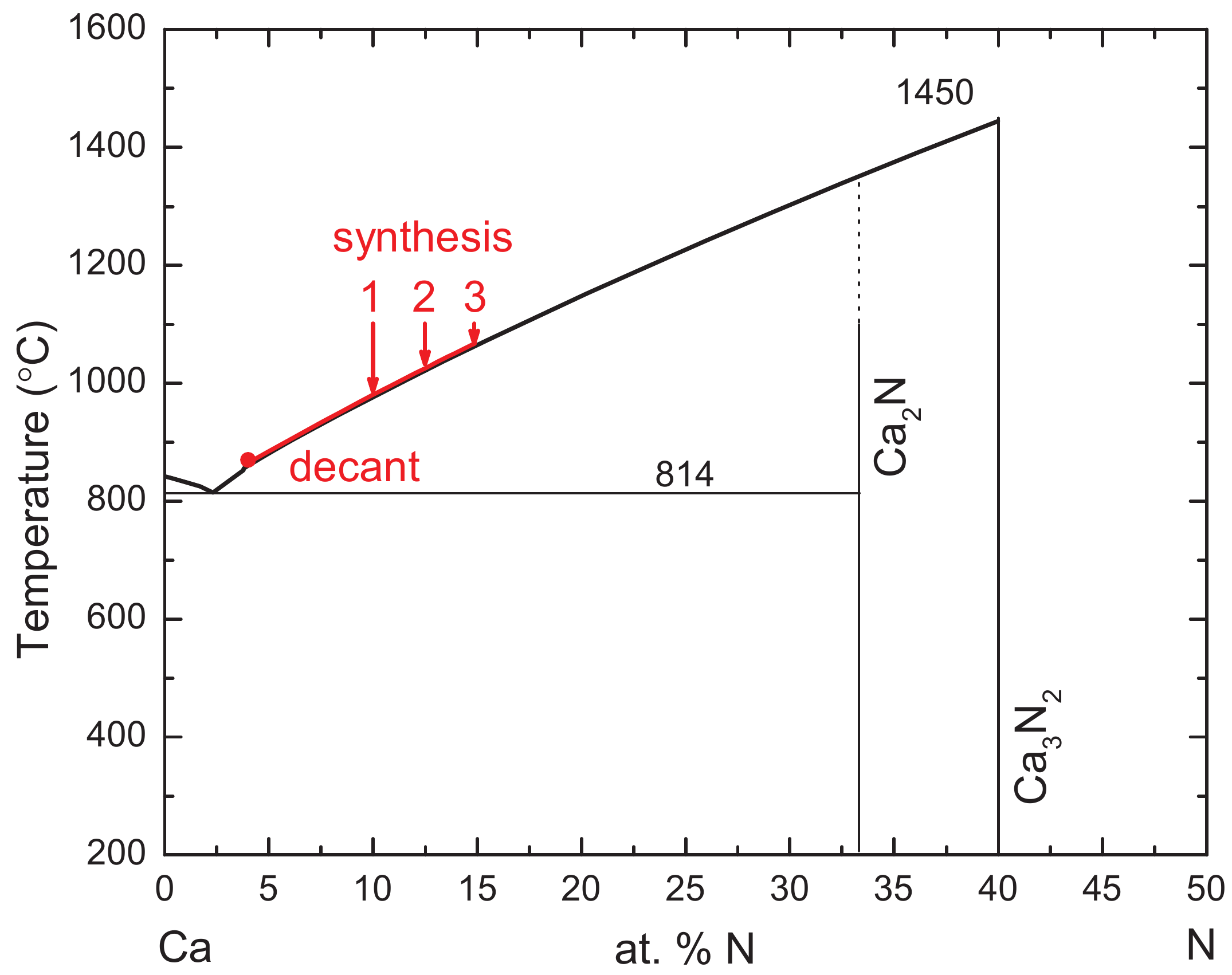}
\caption{Revised Ca-N phase diagram based on\,\cite{Itkin1990,Hohn2009}. The starting compositions (1, 2, and 3) and temperature profile are shown with the red line. 
Transformations between Ca$_2$N, Ca$_3$N$_2$, and $\beta$-Ca$_3$N$_2$ (not shown) depend on temperature and (nitrogen) pressure as reported by H\"ohn \textit{et al.}\,\cite{Hohn2009}. 
}
\label{CaN_phasediagramm}
\end{figure}

\subsection{Ca$_3$N$_2$ and the subnitride Ca$_2$N}

\begin{figure}
\center
\includegraphics[width=0.9\textwidth]{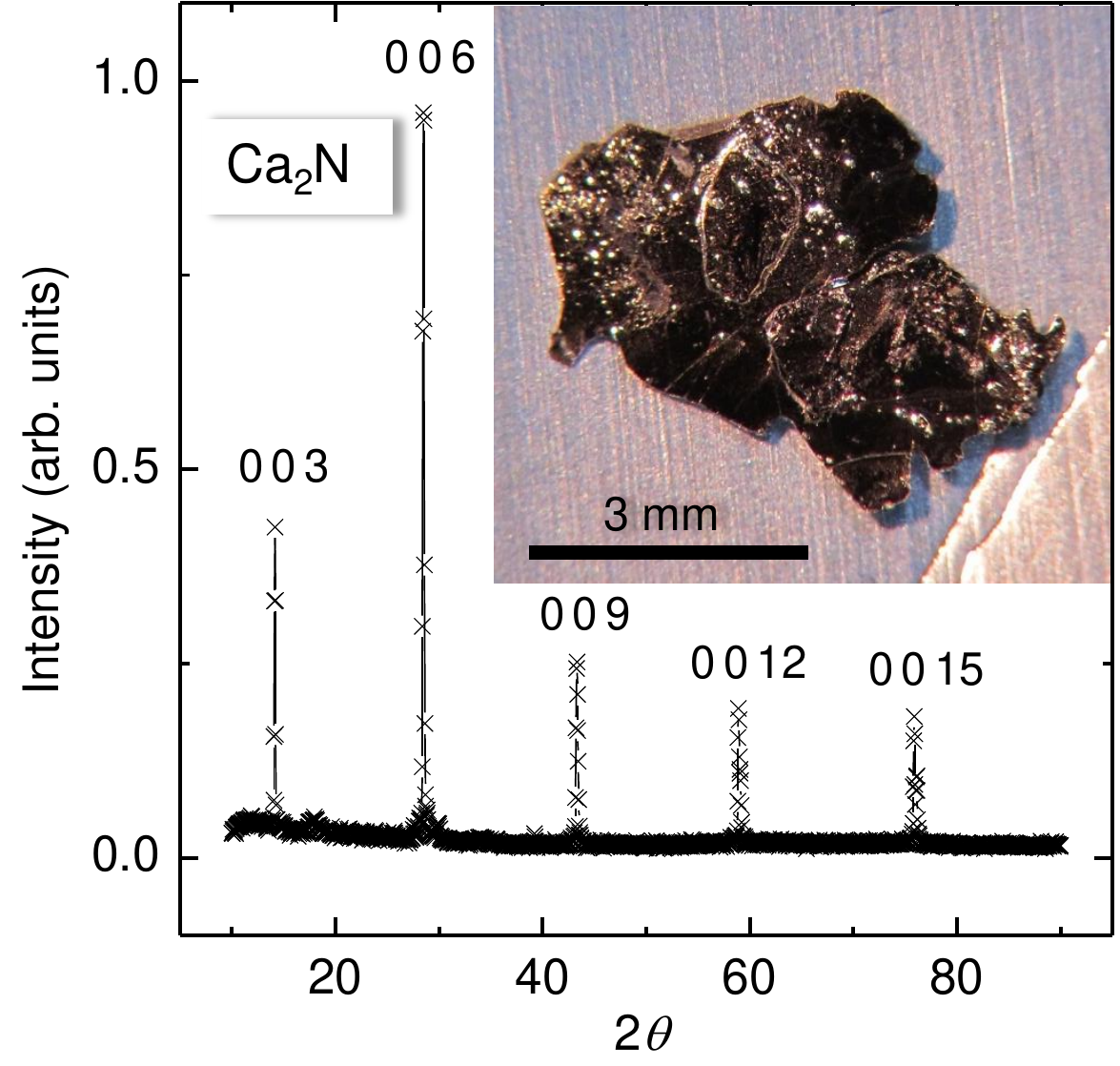}
\caption{Single crystal of Ca$_2$N with a thickness of $\approx$50\,$\mu$m obtained from a Ca$_{85}$N$_{15}$ melt. The X-Ray diffraction pattern was recorded under nitrogen atmosphere on a single crystal with the surface normal of the plate-like sample parallel to the scattering vector.}
\label{Ca2N}
\end{figure}

The high solubility of nitrogen in liquid Ca is demonstrated in the binary alloy phase diagrams (Fig.\,\ref{CaN_phasediagramm} after\,\cite{Itkin1990}).
The phase transformations between Ca$_2$N, Ca$_3$N$_2$, and a third binary, $\beta$-Ca$_3$N$_2$ (not shown in Fig.\,\ref{CaN_phasediagramm}), are reported to depend on temperature and (nitrogen) pressure\,\cite{Hohn2009}.
It should be noted that Ca$_2$N is not included in the early binary phase diagram\,\cite{Itkin1990}. 
In full analogy to the nitride growth from Li-rich flux described in the previous section, the first step to grow nitrides out of a Ca-rich flux was an attempt to grow either Ca$_3$N$_2$ or Ca$_2$N single crystals. 

The starting materials used were Ca and Ca$_3$N$_2$. 
They were mixed in molar ratios of 15:1, 11:1, and 25:3 so as to give initial melt stoichiometries of Ca$_{90}$N$_{10}$, Ca$_{87.5}$N$_{12.5}$, and Ca$_{85}$N$_{15}$, respectively.
The mixtures, each with a total mass of roughly 2.5\,g, were packed in a three-cap Ta and Nb crucibles, heated from room temperature to $T = 1100^\circ$C over 5.5\,h, slowly cooled to 880$^\circ$C over 38 to 122\,h, and finally decanted to separate the single crystals from the excess flux. 
The Nb crucibles were found to be better than Ta crucibles which became brittle at temperatures above 1000$^\circ$C and/or high N concentrations. 

We obtained thin, plate-like single crystals of Ca$_2$N with lateral dimensions of up to 5\,mm and thickness of typically $50\,\mu$m (Fig.\,\ref{Ca2N}). 
The initial composition of the melt had only minor effects on the phase formation and crystallization. Only a small tendency towards larger and thicker Ca$_2$N single crystals was found for Ca:Ca$_3$N$_2$ = 25:3\,$\equiv$\,Ca$_{85}$N$_{15}$.
No traces of the starting material Ca$_3$N$_2$ were found.
From these data we can infer that there is primary solidification of Ca$_2$N for each of these melts with the liquidus line existing below $1100^\circ$C, even for Ca$_{85}$N$_{15}$.

Ca$_2$N is highly air-sensitive\,\cite{Gregory2000} and the single crystals grown from Ca-rich flux show the formation of a white-colored surface layer within a few minutes on air. However, it takes several hours for the samples to completely decompose indicating that the surface layer is passivating to a certain degree. 
The Ca$_2$N single crystals are too malleable to grind them into a fine powder. Therefore, X-Ray diffraction was performed on single crystals with the surface normal of the plate-like samples parallel to the scattering vector. 
The main peaks in the diffraction pattern (Fig.\,\ref{Ca2N}) can be indexed based on $0\,0\,3l$ reflections as expected for Ca$_2$N (according to the reflection conditions for space group $R\,\bar{3}\,mf$ with hexagonal axes).
 
The electrical resistance of Ca$_2$N at room temperature was estimated using a standard multimeter in an Ar filled glove box. 
Contacting the single crystals with two tungsten needles, which were gently stuck into the sample, revealed an electrical resistance too small to be quantified (less than 1\,$\Omega$ corresponding to a resistivity of a few m$\Omega$cm).
However, several attempts to contact samples in four point geometry using silver paint, silver epoxy or carbon paste led to contact resistances of several hundred Ohms and above. 
Curing the contacts or reducing the contact resistance by annealing or applying electric currents failed (the samples were always handled in Ar or nitrogen filled glove boxes). 
An extreme sensitivity to solvents present in standard contact pastes might be at the origin of the enhanced resistance.
For polycrystalline Ca$_2$N differing values of $\rho = 0.16$\,$\Omega$cm\,\cite{Hohn2009} and $\rho = 63$\,$\Omega$cm\,\cite{Gregory2000} were reported (measured at room temperature).
It is reasonable to assume that the electrical transport of this '2D "excess electron" compound'\,\cite{Gregory2000} is highly anisotropic.
A further optimization of the growth procedure and the development of suitable electric contacts may be necessary to study the electric transport in detail and to evaluate possible similarities to other 2 dimensional electron systems like graphene. 

\subsection{Non-nitrides}

\begin{figure}
\center
\includegraphics[width=0.8\textwidth]{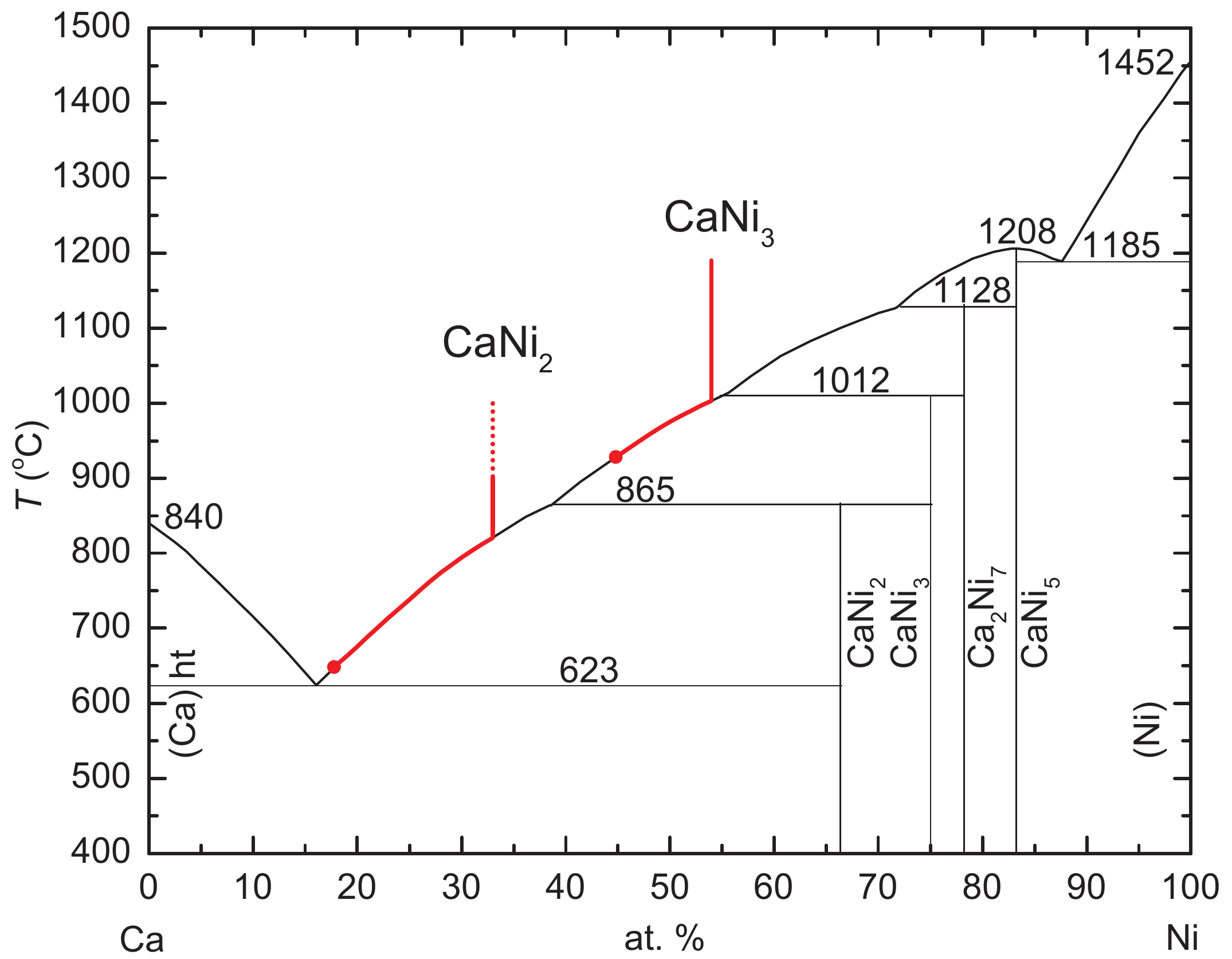}
\caption{Ca-Ni phase diagram after\,\cite{Okamoto1990CaNi}. 
Starting compositions, temperature profile and composition of the liquid phase are shown by the red lines (the dotted line represents rapid cooling).}
\label{phase_cani}
\end{figure}  

\subsubsection{CaNi$_2$ and CaNi$_3$}

\begin{figure}
\center
\includegraphics[width=0.85\textwidth]{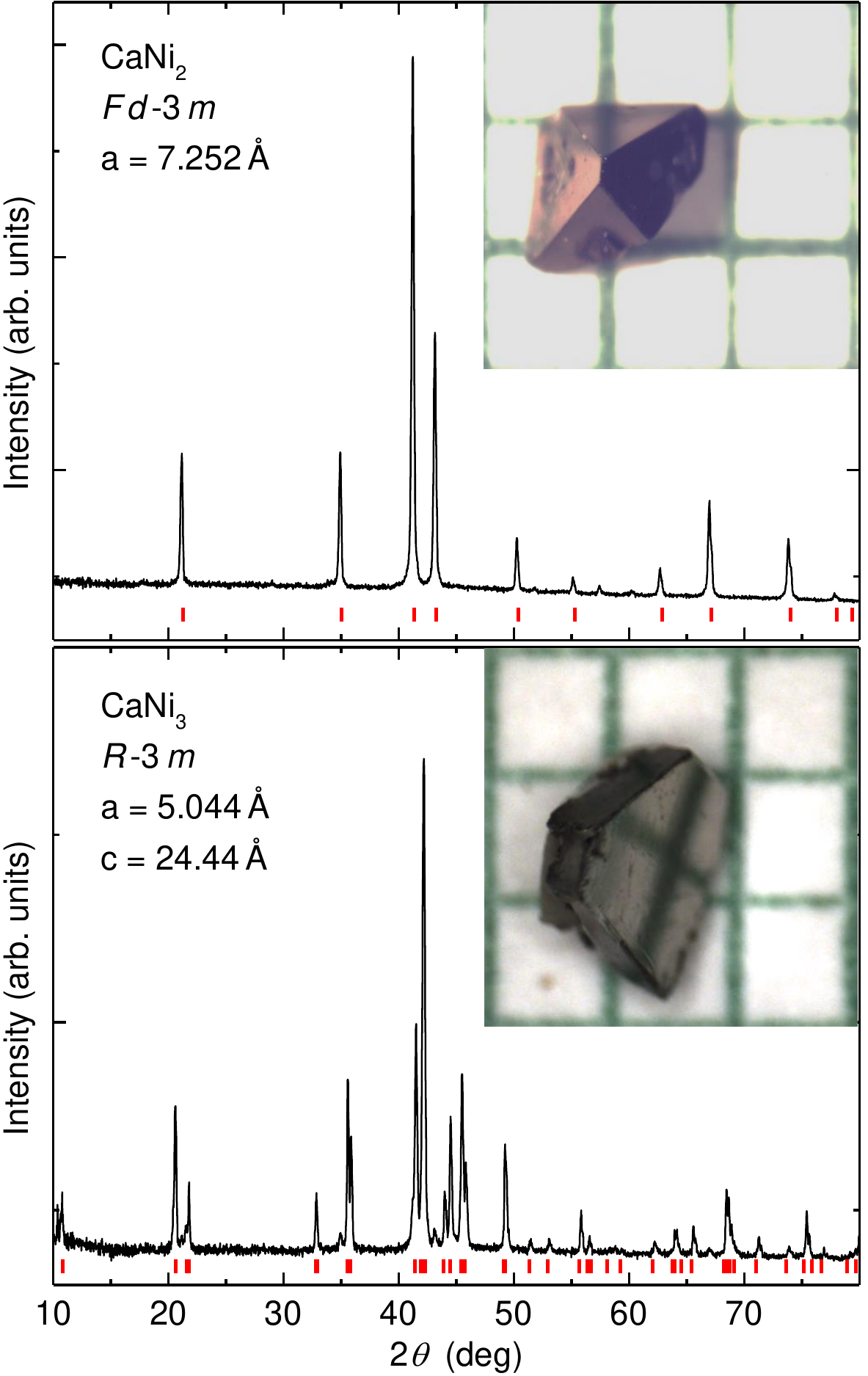}
\caption{X-Ray powder diffraction pattern of CaNi$_2$ and CaNi$_3$ measured on ground single crystals (in air). 
Representative single crystals are shown on a millimeter grid.
}
\label{cani2_cani3_diff_sc}
\end{figure}

Single crystals of CaNi$_2$ and CaNi$_3$ with dimensions of several millimeters were grown from a Ca-rich self-flux (Fig.\,\ref{cani2_cani3_diff_sc}).
Both compounds were found to be metals in close vicinity to a ferromagnetic instability. 
CaNi$_2$ crystallizes in the cubic MgCu$_2$ structure type (space group $F\,\bar{d}\,3\,m$)\,\cite{Buschow1974}. The lattice parameter was found to be $a = 7.252$\,\AA.
CaNi$_3$ crystallizes in the trigonal PuNi$_3$ structure type (space group $R\,\bar{3}\,m$)\,\cite{Buschow1974}. The lattice parameters were found to be $a = 5.044$\,\AA~and $c = 24.44$\,\AA.
Details of the physical measurements and discussion of temperature-dependent magnetization, electrical resistivity, and specific heat have been published separately\,\cite{Jesche2012}.
The finding of almost itinerant ferromagnetism in these two Ca-based compounds motivated several attempts to tune or modify the magnetic properties by substituting Yb, Y, and La for Ca.

Motivated by the binary phase diagram (Fig.\,\ref{phase_cani}) the starting materials for CaNi$_2$ were mixed in a molar ratio of Ca:Ni = 67:33.
The mixtures with a total mass of roughly 2.5\,g were packed in three-cap Ta crucibles, heated from room temperature to $T = 1000^\circ$C over 5\,h, cooled to 900$^\circ$C within 1\,h, slowly cooled to 650$^\circ$C over 50\,h, and finally decanted. 
The starting materials for CaNi$_3$ were mixed in a molar ratio of Ca:Ni = 46:54.
The mixtures with a total mass of roughly 2.5\,g were packed in three-cap Ta crucibles, heated from room temperature to $T = 1190^\circ$C over 6\,h, held for 1/2\,h, cooled to 910$^\circ$C over 32\,h and finally decanted. 
CaNi$_2$ single crystals of octahedral habit with dimensions up to 3\,mm were could be obtained (inset Fig.\,\ref{cani2_cani3_diff_sc}). The CaNi$_3$ single crystals show a platelike habit with lateral dimensions of up to 3\,mm and a thickness of 0.5\,mm (inset Fig.\,\ref{cani2_cani3_diff_sc}).

\subsubsection{YbNi$_2$}
\begin{figure}
\center
\includegraphics[width=0.9\textwidth]{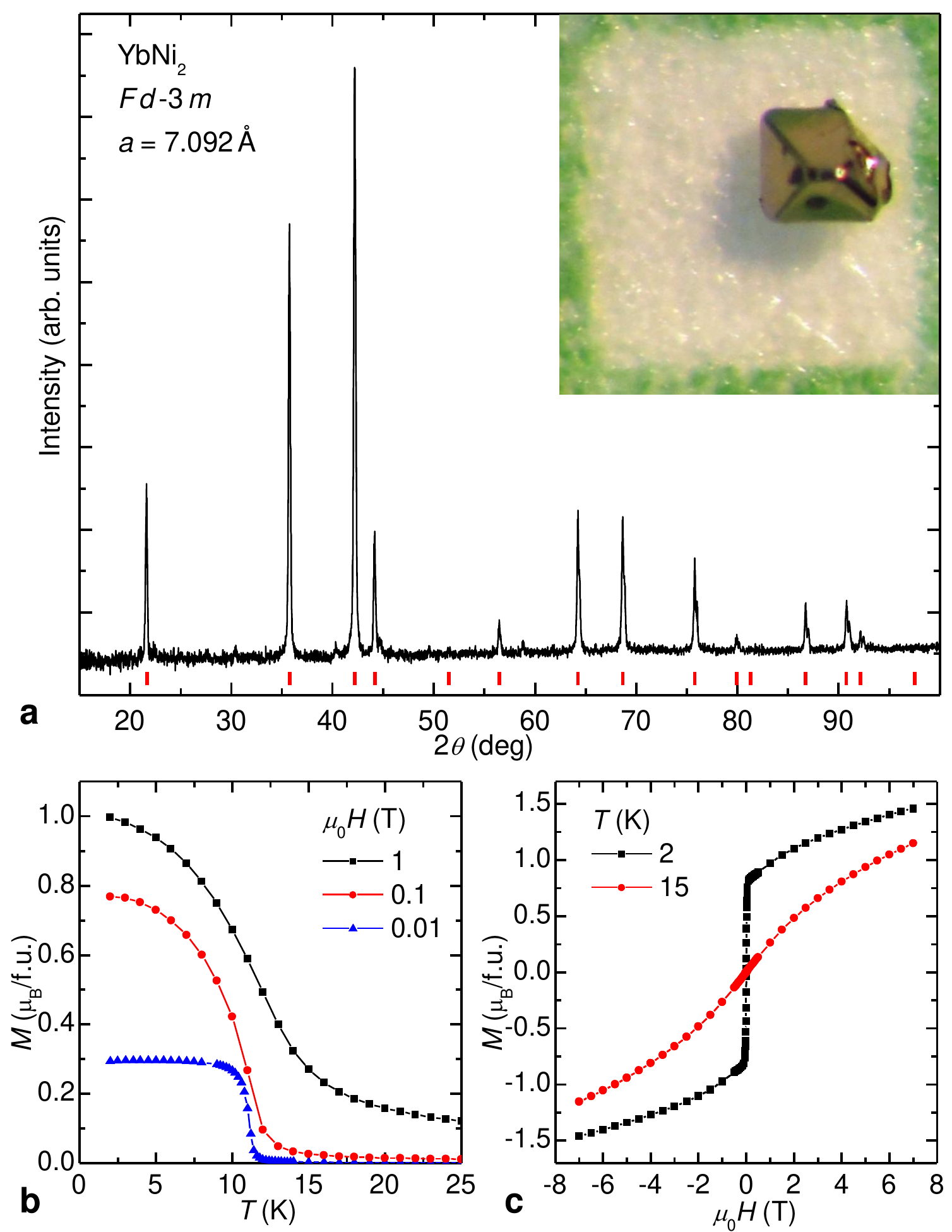}
\caption{a, X-Ray powder diffraction pattern of YbNi$_2$ (collected in air). The theoretical peak positions are marked by red bars.
The inset shows a single crystal of YbNi$_2$ on a millimeter grid.
b, The temperature dependent magnetization of YbNi$_2$ increases step-like for cooling below $T = 12$\,K indicating ferromagnetic ordering. 
c, The isothermal magnetization increases rapidly in small applied fields consistent with the proposed ferromagnetic ordering. 
Polycrystalline samples, synthesized without Ca, show similar behavior\,\cite{Rojas2012}, indicating the absence of significant amounts of Ca impurities in the flux grown single crystals. 
}
\label{YbNi2}
\end{figure}
Single crystals of YbNi$_2$ (isostructural to CaNi$_2$) were grown out of a Ca-rich, ternary melt.
The starting materials were mixed in a molar ratio of Ca:Yb:Ni = 5:1:2 $\equiv$\,Ca$_{62.5}$Yb$_{12.5}$Ni$_{25}$.
The mixture with a total mass of roughly 2.5\,g was packed in a three-cap Ta crucible, heated from room temperature to $T = 1200^\circ$C over 6\,h, held for 0.5\,h, slowly cooled to 900$^\circ$C over 36\,h, and finally decanted.
Single crystals of dimensions up to 0.4\,mm were obtained. The size of the crystals can be likely increased by optimizing the growth procedure. 

Energy dispersive X-Ray analysis, performed on clean, as grown surfaces, indicate a complete absence of Ca in the YbNi$_2$ matrix. This indicates, that, even though CaNi$_2$ and YbNi$_2$ are isostructural,  YbNi$_2$ forms out of a Ca melt with no detectable incorporation of Ca.
XRD measurements on ground single crystals (Fig.\,\ref{YbNi2}a) revealed lattice parameter of $a = 7.092$\,\AA.
This in good agreement with the reported values for YbNi$_2$ ($a = 7.0998(3)$\,\AA\,\cite{Rojas2012}, $a = 7.094$\,\AA\,\cite{Buschow1972}) and significantly smaller than the values found for CaNi$_2$ ($a = 7.252$\,\AA\,\cite{Jesche2012}). 

Magnetization measurements were performed on clusters of small YbNi$_2$ single crystals. The temperature dependent magnetization data, in Bohr magneton per formula unit ($\mu_{\rm B}$/f.u.), are shown in Fig.\,\ref{YbNi2}b. The magnetization in small applied fields increases step-like upon cooling below $T = 12$\,K in accordance with the proposed ferromagnetic ordering\,\cite{Rojas2012}. 
The width of the transition is certainly not larger than in polycrystalline samples (compare with Fig.\,2 in\,\cite{Rojas2012}) in accordance with good quality samples. 
Isothermal measurements at $T = 2$\,K reveal a strong increase of the magnetization in small applied fields but negligible hysteresis (Fig.\,\ref{YbNi2}c). 
The isothermal magnetization at $T = 2$\,K keeps increasing up to the largest applied field of 7\,T. 
A lower limit of 1.5\,$\mu_{\rm B}$/f.u. can be set for the saturation magnetization. 
Similar field-dependence and magnetization values are observed in polycrystalline material\,\cite{Rojas2012}.
The temperature-dependent inverse magnetic susceptibility (not shown) follows a Curie-Weiss behavior for $T > 50$\,K. 
The effective moment is found to 4.2(2)\,$\mu_{\rm B}$/f.u. in fair agreement with the expectation value for Yb$^{3+}$ (4.54\,$\mu_{\rm B}$). An antiferromagnetic Weiss temperature of $\Theta_{\rm W} = -13(4)$\,K is observed.

\subsubsection{Y$_2$Ni$_7$}

\begin{figure}
\center
\includegraphics[width=0.9\textwidth]{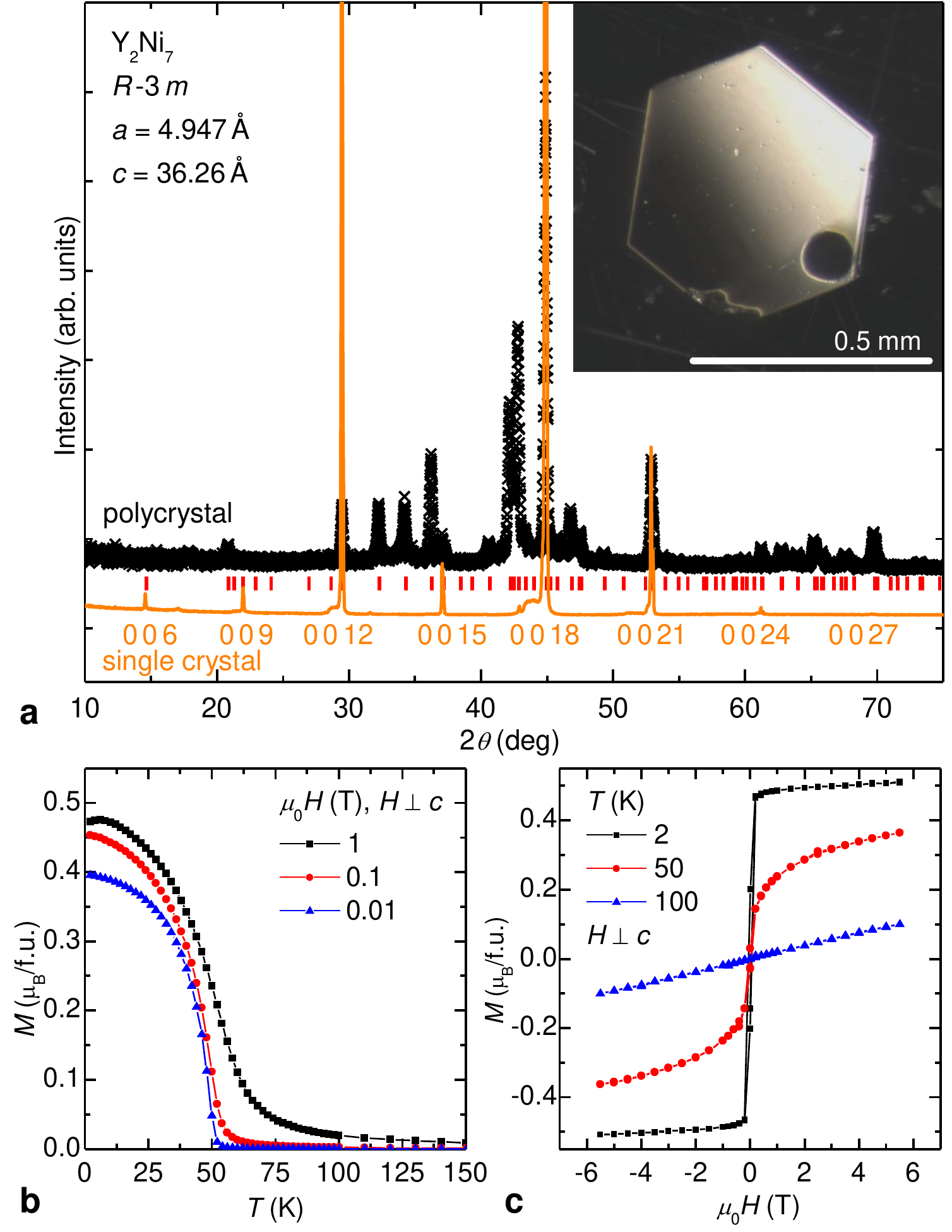}
\caption{a, X-Ray diffraction pattern of Y$_2$Ni$_7$ measured (in air) on ground single crystals (black crosses) and on a single crystal mounted with the plate-like surface laid flat against the sample holder, i.e. with the surface normal of the sample parallel to the scattering vector (solid, orange line).
The theoretical peak positions are marked by red bars.
The peaks observed for the single crystal can be indexed based on the $0\,0\,3l$ reflections and the obtained $c$-axis lattice parameters are in good agreement with the literature data.
The inset shows a single crystal of Y$_2$Ni$_7$ with scale bar.
b, The step-like increase in the temperature-dependent magnetization of Y$_2$Ni$_7$ for cooling below $T = 55$\,K and c, the rapid increase in small applied fields indicate ferromagnetic ordering. 
Polycrystalline samples, synthesized without Ca, show similar behavior\,\cite{Buschow1984}, indicating the absence of significant amounts of Ca impurities in the flux grown single crystals. 
}
\label{YNi}
\end{figure} 

In order to further explore the potential of Ca-Ni-$R$ ($R$ = rare earth element) melts we shifted to $R$ = Y.
An initial melt with the molar ratios Ca\,:\,Y\,:\,Ni = 41.9\,:\,4.5\,:\,53.6 mainly led to the formation of CaNi$_3$ single crystals possibly with some Y substituted for Ca. However, X-Ray powder diffraction pattern, recorded on several batches of crushed single crystals, indicated significant amounts of YNi$_5$ even though the collected single crystals showed a similar habit. 
If both phases are inter grown within the same single crystal could not be settled unambiguously. 
Therefore, these results are briefly discussed in section\,\ref{sec-fail}.
Increasing the Y concentration by a factor of 2 to Ca\,:\,Y\,:\,Ni = 37.5\,:\,9.0\,:\,53.6 led to the formation of Y$_2$Ni$_7$ (Gd$_2$Co$_7$ structure type). An average Ca concentration of 2(1)\%, close to the resolution limit, was found by energy dispersive X-Ray analysis.
The mixtures were packed in three-cap Ta crucibles, heated from room temperature to $T = 1190^\circ$C over 6\,h, held for 1/2\,h, cooled to T = 910$^\circ$C over 32\,h, and finally decanted to separate the single crystals from the excess flux.
We obtained Y$_2$Ni$_7$ single crystals with lateral dimensions of up to 0.5\,mm and thickness of $\sim0.1$\,mm (Fig.\,\ref{YNi}a).

It is worth mentioning that the synthesis of Y$_2$Ni$_7$ from an Y-rich self-flux is not possible at temperatures below 1200$^\circ$C and that excludes the use of silica ampules (which protect the Ta crucible). In contrast, single crystal growth from Ca-flux can be performed at temperatures below 1200$^\circ$C and that is of great practical importance for the growth and optimization of this compound. 
We are not aware of any previous report on the synthesis of single crystalline Y$_2$Ni$_7$.

XRD measurements on ground Y$_2$Ni$_7$ single crystals (Fig.\,\ref{YNi}a) revealed lattice parameters of $a = 4.947$\,\AA,~and $c = 36.26$\,\AA. 
This in good agreement with the reported values of $a = 4.924$\,\AA~and $c = 36.07$\,\AA\,\cite{Virkar1969} (Gd$_2$Co$_7$ structure type).
X-Ray diffraction was also performed on single crystals with the surface normal of the plate-like samples parallel to the scattering vector (Fig.\,\ref{YNi}a). 
The main peaks in the diffraction pattern can be indexed based on $0\,0\,3l$ reflections as expected for Y$_2$Ni$_7$ (according to the reflection conditions for space group $R\,\bar{3}\,m$ with hexagonal axes). 
Under standard laboratory conditions the samples are not air-sensitive on a timescale of one year.

The temperature dependent magnetization of Y$_2$Ni$_7$ increases strongly upon cooling below $T = 55$\,K (Fig.\,\ref{YNi}b), in accordance with the reported ferromagnetic ordering observed in polycrystalline samples\,\cite{Buschow1984}. 
The onset and the temperature dependence of the increase are very similar to the behavior of polycrystalline Y$_2$Ni$_7$ (compare with Figs.\,1 and 2 in\,\cite{Buschow1984} and \cite{Nishihara1991,Bhattacharyya2011}).
Isothermal magnetization measurements at $T = 2$\,K reveal a strong increase and fast saturation of the magnetization at $0.51\,\mu_{\rm B}$/f.u. (Fig.\,\ref{YNi}c) corresponding to $0.07(1)\,\mu_{\rm B}$/Ni in good agreement with the reported value ($0.08\,\mu_{\rm B}$/Ni). 
The rather small saturation moment is in accordance with the decrease of the saturation moment with decreasing Curie temperature as observed in many Ni-based metallic compounds\,\cite{Buschow1984}.
Furthermore, it has been reported that the size of the ordered moment in Y$_2$Ni$_7$ significantly dependents on Ni vacancies\,\cite{Nakabayashi1992}. 
The moment was found to increase linearly from $0.033\,\mu_{\rm B}$/Ni in Y$_2$Ni$_7$ to $0.083\,\mu_{\rm B}$/Ni in Y$_2$Ni$_{6.7}$. Considering these results, albeit obtained on polycrystalline material, a slight Ni deficiency might be inferred for the single crystals reported here.

The similarities to polycrystalline samples, which were synthesized by arc melting and annealing of Y and Ni\,\cite{Buschow1984}, indicates a negligible effect of possible Ca contaminations or substitutions on the magnetic properties of Y$_2$Ni$_7$ or, more likely, the absence of Ca in the single crystals.
Similar to the case of YbNi$_2$, the replacement of Ca by a rare earth element seams to be energetically highly favorable in these crystal structures as soon as the rare earth concentration in the melt exceeds a certain threshold. 
This is even more surprising considering the similar lattice parameters of Ca$_2$Ni$_7$\,\cite{Buschow1974} and Y$_2$Ni$_7$\,\cite{Buschow1984} with $\Delta a/a = 1.2\%$ and $\Delta c /c = -0.3\%$ ($\Delta a = a_{\rm Ca}-a_{\rm Y}$ and $\Delta c = c_{\rm Ca}-c_{\rm Y}$). 
The different valence states, predominantly Ca$^{2+}$ vs. Y$^{3+}$, are likely the dominating factor for the observed phase formation. 

\subsubsection{LaNi$_5$}
\begin{figure}
\center
\includegraphics[width=0.9\textwidth]{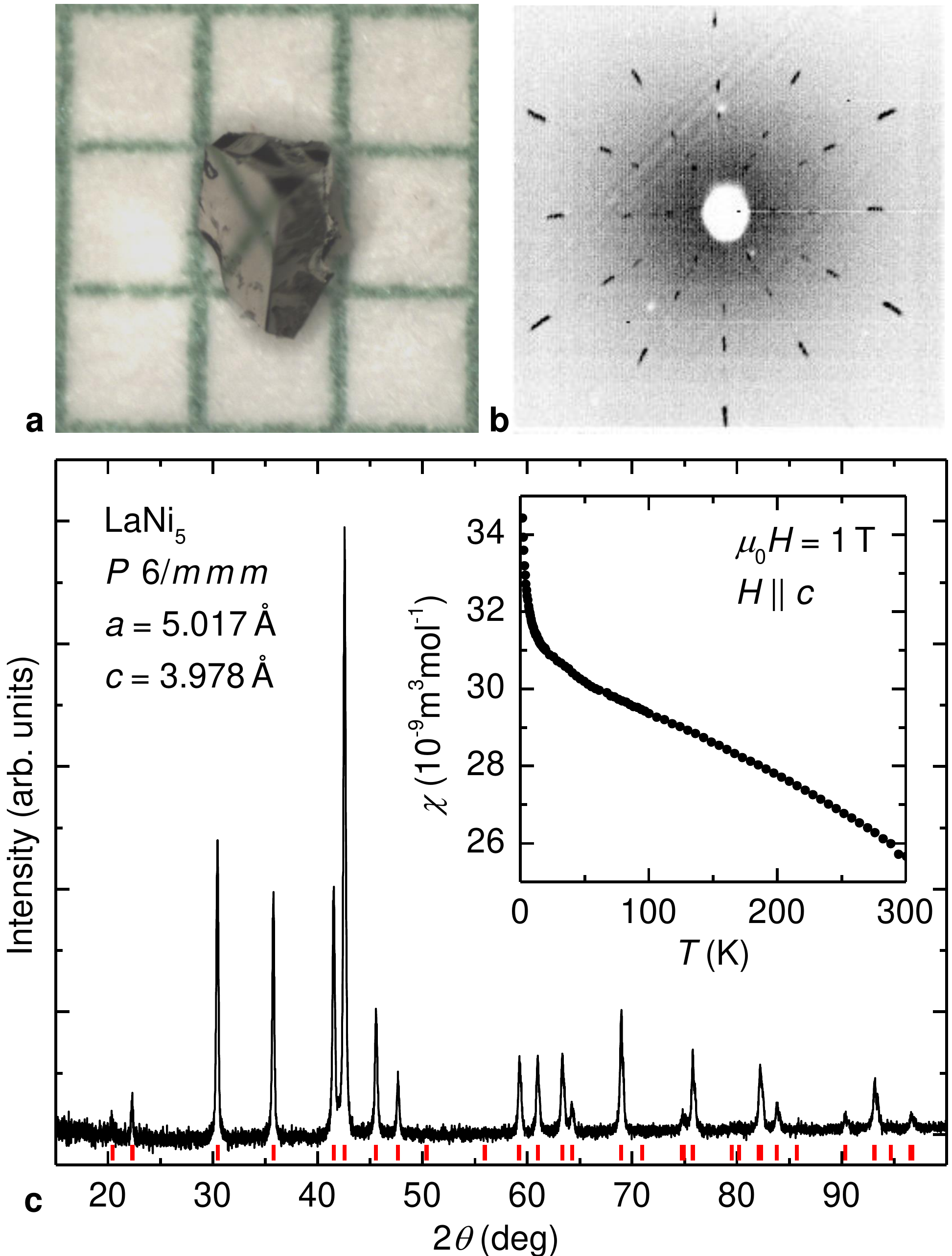}
\caption{ a, LaNi$_{5}$ single crystal on a millimeter grid and b, corresponding Laue-back-reflection pattern. 
c, X-Ray powder diffraction pattern of LaNi$_5$ measured on ground single crystals (in air). The theoretical peak positions for the reported structure are marked by red bars.
The inset shows the temperature-dependent magnetic susceptibility $\chi = M/H$. 
}
\label{LaNi}
\end{figure} 

Exploration of the Ca-Ni-La ternary melt lead to the growth of LaNi$_5$.
An initial melt with the molar ratios Ca\,:\,La\,:\,Ni = 41.9\,:\,4.5\,:\,53.6  led to the formation of LaNi$_5$. 
The elements were packed in three-cap Ta crucibles, heated from room temperature to $T = 1190^\circ$C over 6\,h, held for 1/2\,h, cooled to T = 910$^\circ$C over 32\,h, and finally decanted to separate the single crystals from the excess flux.
The LaNi$_5$ single crystals had an isometric habit (in accordance with a rather small $c/a$ ratio) with typical dimensions of 1\,mm.
Laue-back-reflection pattern show the six-fold rotation symmetry along the crystallographic $c$-direction (Fig.\,\ref{LaNi}b).
XRD measurements on ground LaNi$_5$ single crystals (Fig.\,\ref{LaNi}c) revealed lattice parameters of $a = 5.017$\,\AA~and $c = 3.978$\,\AA.
This is in good agreement with the reported values of $a = 5.013$\,\AA~and $c = 3.984$\,\AA\,\cite{Wernick1959}.
An average Ca concentration of 2(1)\,at.\%, close to the resolution limit, was found by energy dispersive X-Ray analysis.
Increasing the La concentration by a factor of 2 to Ca\,:\,La\,:\,Ni = 37.5\,:\,9.0\,:\,53.6 led to the formation LaNi$_{x}$ with $x \approx 4$ possibly with some Ca substituted for La. 
However, we found no unique solution to index the X-Ray powder diffraction pattern based on one structure. Therefore, these results are briefly discussed in section\,\ref{sec-fail}.

The magnetic susceptibility at room temperature was found to be $\chi = 26\cdot10^{-9}$\,m$^3$mol$^{-1}$ ($H \parallel c$, $\mu_0H = 1$\,T) in good agreement with the reported value of $\chi = 23\cdot10^{-9}$\,m$^3$mol$^{-1}$ obtained on polycrystalline material\,\cite{Burzo2000}. 
$\chi(T)$ shows a roughly linear increase by $\sim20$\% upon cooling to $T = 20$\,K (inset Fig.\,\ref{LaNi}c).
A similar increase was also observed in polycrystalline material for cooling down to $\sim100$\,K\,\cite{Burzo2000}.
A further, enhanced increase upon cooling below $\sim20$\,K (Curie tail) can be attributed to a small amount of local moment bearing magnetic impurities.

\section{Ambiguous and failed attempts}\label{sec-fail}
In addition to the crystal growths detailed above, there have been growth attempts that have led to either ambiguous phases or failure to yield sufficiently large samples by the decanting temperature. The classification of the results as being ambiguous or failed is by no means meant to be a final statement.
Very small changes in the composition of the ternary, quaternary or quinary melts and/or variations in the temperature profile can have dramatic effects on the phase formation.
The intention is rather to give a comprehensive overview of growth attempts for which the result was:
\begin{itemize}
\item dramatically different from the goal of the synthesis,
\item not identifiable,
\item a total spin (decanting above the liquidus temperature), 
\item polycrystalline material which can be easily obtained else-wise.
\end{itemize}

These growths include attempts to incorporate Ag, As, Au, B, Cr, Cu, Mg, Sc, Tb, Ti, and V into Li or Ca based melts.
Two compounds, ({\bf Ca},Y)Ni$_3$ and LaNi$_x$, are presented in more detail at the beginning of this section whereas the other growth attempts are briefly described in tabular form below (Tab.\,\ref{failedgrowth}).

\begin{figure}
\center
\includegraphics[width=0.9\textwidth]{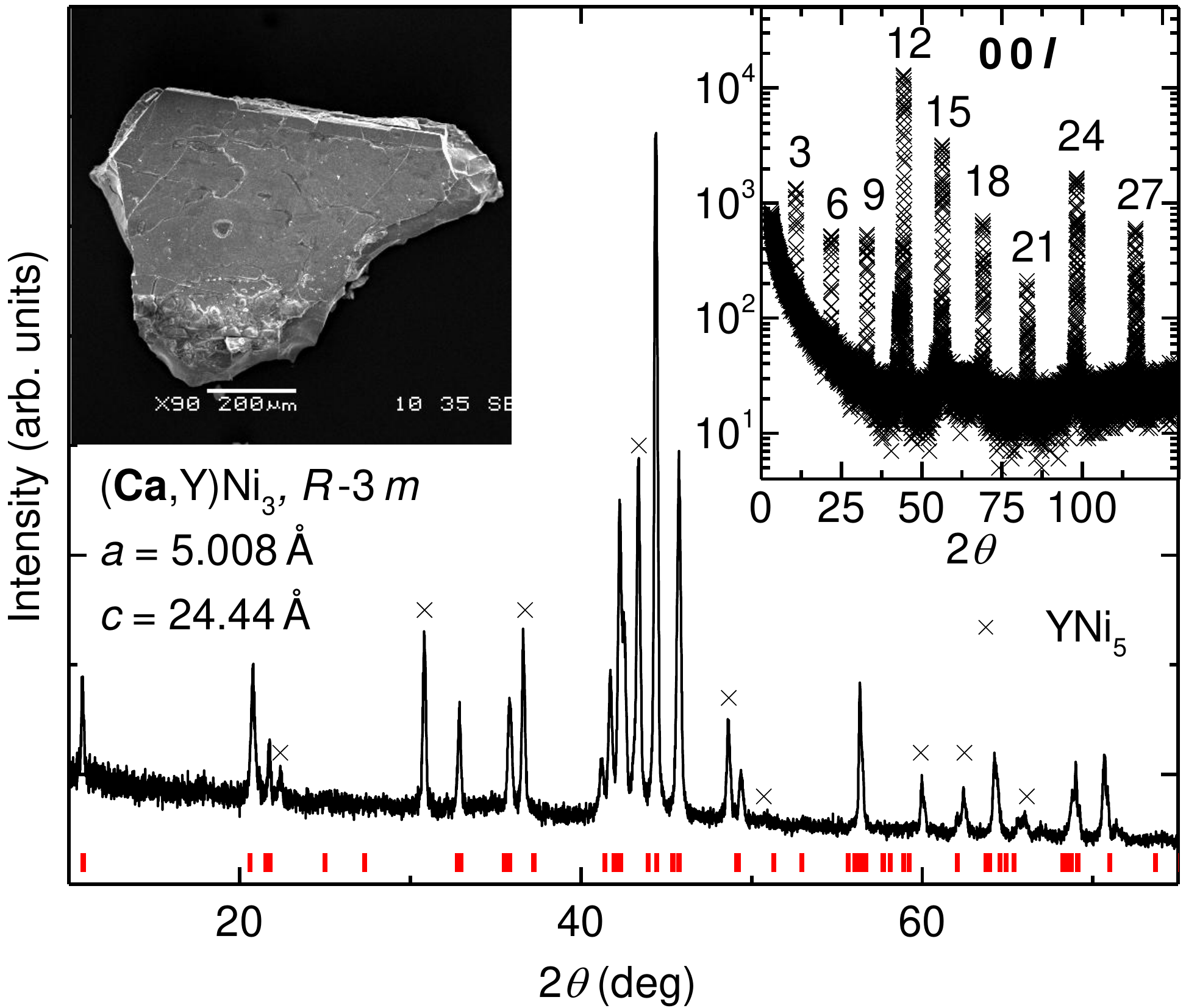}
\caption{
X-Ray powder diffraction pattern of ({\bf Ca},Y)Ni$_3$ measured on ground single crystals (in air).
An electron micrograph of an as-grown sample is shown as inset to the left.
The right hand inset shows the diffraction pattern obtained on this single crystal which was mounted with the plate-like surface laid flat against the sample holder, i.e. with the surface normal of the sample parallel to the scattering vector.
The peaks observed for the single crystal can be indexed based on the $0\,0\,3l$ reflections and the obtained $c$-axis lattice parameters are in good agreement with the literature data (the number given in the plot is the $l$ Miller index).
}
\label{cayni3}
\end{figure}

({\bf Ca},Y)Ni$_3$:
An initial melt with the molar ratios Ca\,:\,Y\,:\,Ni = 41.9\,:\,4.5\,:\,53.6  led to the formation of CaNi$_3$ possibly with some Y substituted for Ca. Energy dispersive X-Ray analysis on four samples (on 17 spots in total) revealed an Y concentration of 4(1)\,at.\% and a Ca concentration of 22(4)\,at.\% corresponding to (Ca$_{0.85}$Y$_{0.15}$)Ni$_3$.
Due to the large uncertainty in the Y concentration, the compound is denoted as ({\bf Ca},Y)Ni$_3$.
The mixtures were packed in three-cap Ta crucibles, heated from room temperature to $T = 1190^\circ$C over 6\,h, held for 1/2\,h, cooled to T = 910$^\circ$C over 32\,h, and finally decanted to separate the single crystals from the excess flux.
We obtained single crystals with lateral dimensions of up to 1\,mm and thickness of $\sim0.2$\,mm for ({\bf Ca},Y)Ni$_3$ (inset Fig.\,\ref{cayni3}). 
XRD measurements on ground single crystals (Fig.\,\ref{cayni3}) can be indexed assuming a mixture of ({\bf Ca},Y)Ni$_3$ and (Ca,Y)Ni$_5$. However, we can not exclude the presence of a possible superstructure\,\cite{Ozaki2007}.
The ({\bf Ca},Y)Ni$_3$ lattice parameters of $a = 5.008$\,\AA,~and $c = 24.44$\,\AA~are close to the values of $a = 5.044$\,\AA~and $c = 24.44$\,\AA~that were found in undoped CaNi$_3$ which was synthesized by the same growth procedure\,\cite{Jesche2012}.
X-Ray diffraction was also performed on single crystals with the surface normal of the plate-like samples parallel to the scattering vector (inset Fig.\,\ref{cayni3}, the given number is the Miller index $l$).
All peaks in the diffraction pattern can be indexed based on $0\,0\,3l$ reflections as expected for CaNi$_3$ (according to the reflection conditions for space group $R\,\bar{3}\,m$ with hexagonal axes).

The magnetic properties of ({\bf Ca},Y)Ni$_3$ are indistinguishable from pure CaNi$_3$ \cite{Jesche2012} within the error of the measurements.

\begin{figure}
\center
\includegraphics[width=0.9\textwidth]{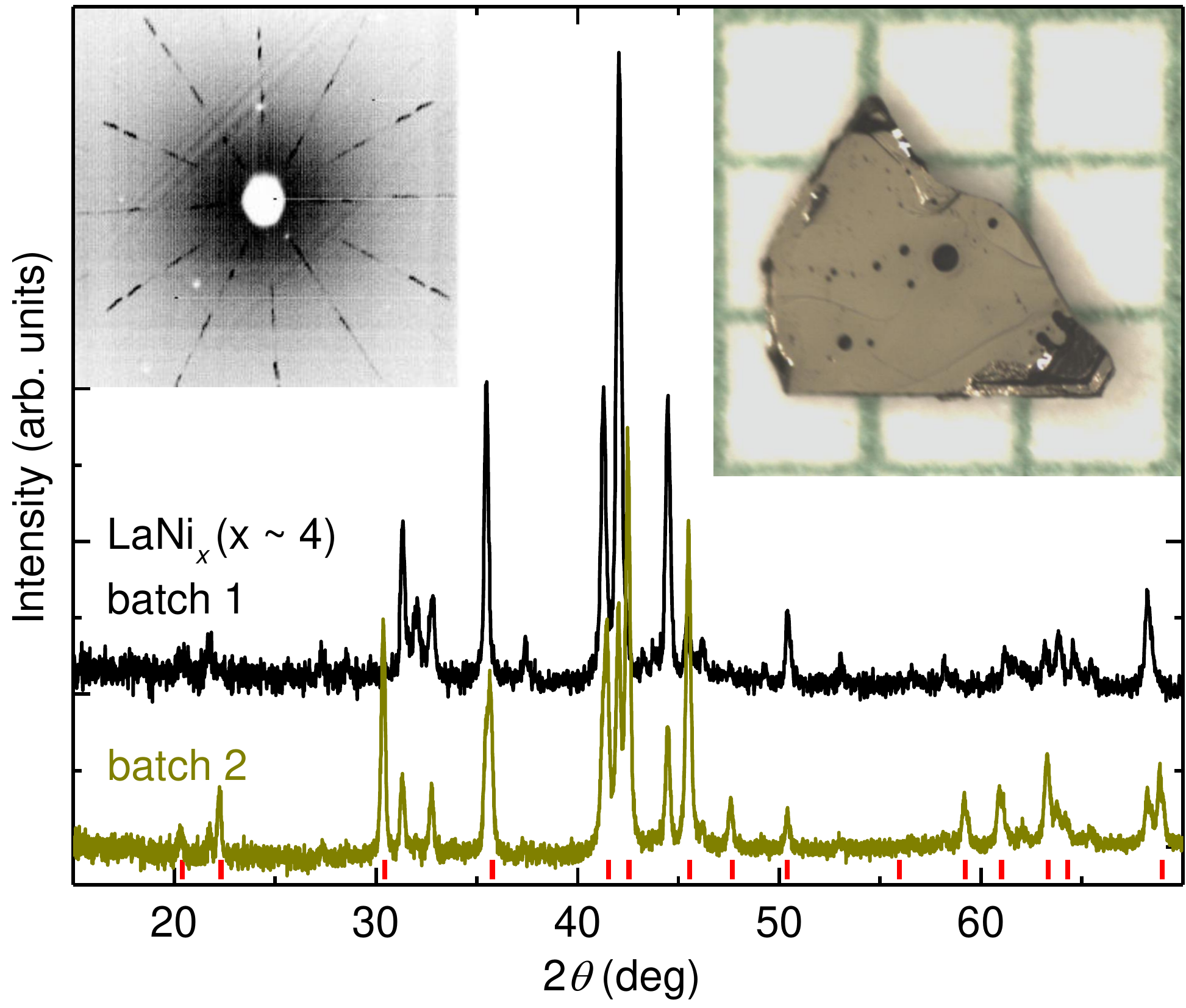}
\caption{ 
Two X-Ray diffraction pattern of LaNi$_x$ measured on ground single crystals (in air).
The theoretical peak positions for LaNi$_5$ are shown by red bars. The right hand inset shows an as-grown single crystal.
A corresponding Laue-back-reflection pattern is shown as inset to the left.
}
\label{lanix}
\end{figure}

LaNi$_x$:
An initial melt with the molar ratios Ca\,:\,La\,:\,Ni = 37.5\,:\,9.0\,:\,53.6 led to the formation of single crystalline LaNi$_{x}$ with $x \approx 4$ possibly with some Ca substituted for La.  
We obtained plate-like single crystals with lateral dimensions of up to 2\,mm and thickness of $\sim0.4$\,mm (Fig.\,\ref{lanix}).
Ca concentrations of 3(1)\,at.\% and Ni concentrations of $\sim80$\,at.\% were estimated by energy dispersive X-Ray analysis.
The majority of X-Ray powder diffraction peaks can be indexed based on the structure of LaNi$_3$ or La$_5$Ni$_{19}$. However, the solution is not unique and additional peaks appear. A measurement on a second batch of ground single crystals also indicated the presence of a significant amount of LaNi$_5$.
Furthermore, X-Ray diffraction on single crystal with the surface normal of the plate-like samples parallel to the scattering vector showed well defined Bragg peaks at low angles indicating a periodicity of 50\,\AA~along the $c$-axis. 
It is known that several La-Ni compounds can be considered as varying stacking sequence build from La$_2$Ni$_4$ and LaNi$_5$ block layers\,\cite{Yamamoto1997}. 
In fact, long-range superstructures with periodicities of 49\,\AA~and above along the $c$-axis have been found in La-Mg-Ni based alloys\,\cite{Ozaki2007}. Therefore, the formation of superstructures in Ca-substituted LaNi$_x$ and in ({\bf Ca},Y)Ni$_3$ seem not unlikely.
However, a further investigation of this possibility is beyond the scope of this publication.

Further growth attempts are summarized in Tab.\,\ref{failedgrowth}. The goal of the growth is given in column one followed by the molar ratio of the starting materials in column two. 
The temperature profile is given in column three where the first number denotes the time in hours to reach the first temperature set point (starting from between room-temperature and $\sim 200^\circ$C). The last number gives the decanting temperature. 
Column four describes the results of the growths.

\section{Summary}
We demonstrate the use of Li and Ca flux for the single crystal growth of various compounds.
The synthesis of large nitride single crystals from Li- or Ca-rich flux is, to our knowledge, demonstrated for the first time and has several advantages compared to the more common reactions from the elements with N$_2$ gas.

Growth attempts for several Li-transition metal binaries are reported and we present a basic, physical characterization of LiRh as well as magnetic properties of Li$_2$Pd, Li$_2$Pt, Li$_3$Al$_2$, YbNi$_2$, Y$_2$Ni$_7$, and LaNi$_5$.
A surprising discovery was the successful growth of single crystalline rare earth-nickel binaries from Ca flux that were essentially free from Ca incorporations. 

The viability of Li and Ca flux is by no means meant to be restricted to the presented compounds. 
Rather, we hope to have demonstrated the wide range of applications these elements can have for growing various compounds and provide practical, technical solutions to overcome the difficulties associated with the high reactivity of Li and Ca.

\section*{Acknowledgment}
S. L. Bud'ko, P. H\"ohn, R. S. Houk, J. L. Jacobs, and A. Kreyssig are acknowledged for comments and discussions. The authors thank F. Laabs and W. E. Straszheim for assistance with energy dispersive X-Ray analysis.  
This work was supported by the U.S. Department of Energy, Office of Basic Energy Science, Division of Materials Sciences and Engineering (BES-DMSE). The research was performed at the Ames Laboratory. Ames Laboratory is operated for the U.S. Department of Energy by Iowa State University under Contract No. DE-AC02-07CH11358.
The authors would like to acknowledge supplemental support from BES-DMSE for the acquisition of a Inert Gas Atmosphere system that was instrumental in the synthesis and characterization of the materials prepared for this research.

\begin{sidewaystable}
\caption{Ambiguous and failed growth attempts}
\begin{tabular}{lllll}
\hline
\hline
Goal	& Initial composition		& Temperature profile ($^\circ$C)	& Result \\
\hline
Ag$_{16}$Ca$_6$N	& Ca:Ag:Ca$_3$N$_2$ = ~33:32:1	& 5h-1180-0.5h-1180-24h-900	&	total spin  \\  
Ag$_8$Ca$_{19}$N$_7$& ~~~~~~~~"~~~~~~~~ = 131:16:7	& "& Ca$_2$N, brittle Ta  \\
\hline
AlLi$_3$N$_2$	& Li:Al:Li$_3$N=11:1:2	& 4h-900-44h-400 & small amount of AlN poly\\
"   			& ~~~~~~~"~~~~~=10:2:2	& " & AlN poly\\
\hline
BN		& Cu:Li:Li$_3$N:B=87:~48:13:13   			& 6h-1200-1h-1200-60h-900	& Li$_3$BN$_2$, Cu$_3$N, unidentified transparent flakes\\
"		& Pd:Li:Li$_3$N:B=67:144:39:33   			& "	& Li$_3$BN$_2$, Li$_2$Pd\\
\hline
Ca$_{3}$AuN	& Ca:Au:Ca$_3$N$_2$=21:2:1	& 6h-1200-0.3h-1200-36h-900	& total spin\\ 

"	& ~~~~~~~~"~~~~~~~~=21:2:1	& 5.5h-1100-1.5h-920-40h-825	& Ca$_2$N\\ 
"	& ~~~~~~~~"~~~~~~~~=15:2:3	& "	& "\\
"	& ~~~~~~~~"~~~~~~~~=33:2:1	& 6h-1200-32h-880	& "\\
\hline
CaMg$_2$N$_2$ & Ca:Mg:Ca$_3$N$_2$=82.5:5:2.5	& 4h-940-1h-940-50h-840 & Ca$_2$N SC\\
\hline
CaNiN	& Ca:Ni:Ca$_3$N$_2$ = 5:4:1 & 5\,h-1000-1\,h-900-50\,h-650	& Ca$_2$N SC\\
"		&   ~~\,~~~~   " ~~~~~~~ = 9:4:1	& "	& " \\
"		&   ~~\,~~~~   " ~~~~~~~ = 13:6:1	& "	& " \\
"		&   ~~\,~~~~   " ~~~~~~~ = 17:8:1	& "	& " \\
\hline
CeFeAsO & Ca:Ce:Fe:As:Fe$_2$O$_3$=60:3:1:3:1	& 6h-1200-0.5h-1200-40-880 & Ce$_4$As$_3$ SC\\
\hline
CeLi$_2$N$_2$	& Li:Ce:Li$_3$N=5.5:0.5:1	& 4.5h-900-1.5h-750-44h-350 & Li$_3$N SC, Ce\\
\hline
LiAgC$_2$	& Li:Ag:C=18:1:2	& 5h-1000-1h-1000-60h-400 & Li$_2$C$_2$ SC + AgLi poly\\ 
"			& ~~~~~"~~~~=18:2:2			& " 						& " \\
\hline
LiFeAs	& Li:FeAs:As=4:1:1	& 6h-850-50h-550	& LiFeAs poly\\
"		& "  				& 9h-950-50h-550	& "\\
"		& "					& 11h-1050-50h-750	& "\\
\hline
LiMgN			& Li:Mg:Li$_3$N=5:1:1	& 4.5h-900-1.5h-750-100h-350& Li$_3$N SC\\
\hline
Li$_2$(Li$_{0.7}$Sc$_{0.3}$)N 	& Li:Sc:Li$_3$N=5.7:0.3:1& 4.5h-900-1.5h-750-100h-350& Li$_3$N SC\\
Li$_2$(Li$_{0.7}$Ti$_{0.3}$)N 	& Li:Ti:Li$_3$N=~~~~" & " & "\\
Li$_2$(Li$_{0.7}$V$_{0.3}$)N 	& Li:V~:Li$_3$N=~~~~" & " & "\\
Li$_2$(Li$_{0.7}$Cr$_{0.3}$)N 	& Li:Cr:Li$_3$N=~~~~" & " & "\\
\hline
Li$_2$Pt$_3$B	& Li:Pt:B=9:3:1 &4h-1000-1h-1000-20h-700& total spin\\

"		& "				& 4h-1000-1h-1000-1.5h-720-20h-500	& Li$_2$Pt\\ 

" 		&Li:Pt:B=69:15:16 	& 4h-700-4h-700-40h-500 & unidentified SC, isometric habit\\
Li$_2$Pd$_3$B & Li:Pd:B=69:15:15 & " 			& unidentified crystalline material, needle-like habit\\
\hline
TbCu$_2$	& Li:Tb:Cu=92.5:2.5:5	& 4h-900-47h-400 & TbCu$_2$ poly\\
"			& ~~~~\,~"~~~~~=85.0:5.0:10	& "					 & "\\

\hline
\hline
\end{tabular}

Crucible material: Ta. Temperature profile: the first number denotes the time (in hours) to reach the first temperature which is given by the second number in $^\circ$C, the third number denotes the time to reach the next temperature, which is given by number 4, etc.
Abbreviations: SC = single crystal, poly = polycrystalline
\label{failedgrowth}
\end{sidewaystable}


\end{document}